\documentclass[12pt]{article}
\usepackage{amsmath}
\usepackage{graphicx,psfrag,epsf}
\usepackage{enumerate}
\usepackage[round]{natbib}

\usepackage{setspace}
\usepackage{graphicx} 
\usepackage{amssymb}
\usepackage{subcaption}
\usepackage{float}
\usepackage{enumerate}
\usepackage{booktabs}
\usepackage{xcolor}
\usepackage{lineno}
\usepackage{newtxtext}
\usepackage{newtxmath}

\usepackage{authblk}            

\usepackage{url} 

\newcommand{\blind}{0}

\newcommand{\N}{\mathrm{N}}
\newcommand{\C}{\mathcal{C}}
\newcommand{\M}{\mathcal{M}}

\newcommand{\E}{\mathrm{E}}

\newcommand\ind{\protect\mathpalette{\protect\independenT}{\perp}}
\def\independenT#1#2{\mathrel{\rlap{$#1#2$}\mkern2mu{#1#2}}}

\newcommand{\CATE}{\text{CATE}}
\newcommand{\ATE}{\text{ATE}}

\begin{document}

\def\spacingset#1{\renewcommand{\baselinestretch}%
{#1}\small\normalsize} \spacingset{1}


\newcommand{\mytitle}{Estimating heterogeneous effects of continuous
  exposures using Bayesian~tree~ensembles: revisiting the impact of
  abortion rates on crime}

\if0\blind
{
  \title{\bf \mytitle}
  
  

  \author[1]{Spencer Woody\thanks{Corresponding author. Email to
      \texttt{spencer.woody@utexas.edu}}}

  \author[2,1]{Carlos M. Carvalho}

  \author[3]{P. Richard Hahn}

  \author[2,1]{Jared~S.~Murray}

\affil[1]{Department of Statistics and Data Sciences, University~of~Texas~at~Austin}

\affil[2]{Department of Information, Risk, and Operations Management, University~of~Texas~at~Austin}

\affil[3]{School of Mathematical and Statistical Sciences, Arizona
  State University}

  \maketitle
} \fi

\if1\blind
{
  \bigskip
  \bigskip
  \bigskip
  \begin{center}
    {\LARGE\bf \mytitle}


    \bigskip
    \today
    
\end{center}
  \medskip
} \fi


\frenchspacing

\vspace{-0.5cm}
\begin{abstract}
  In estimating the causal effect of a continuous exposure or
  treatment, it is important to control for all confounding factors.
  However, most existing methods require parametric specification for
  how control variables influence the outcome or generalized
  propensity score, and inference on treatment effects is usually
  sensitive to this choice.  Additionally, it is often the goal to
  estimate how the treatment effect varies across observed units.  To
  address this gap, we propose a semiparametric model using Bayesian
  tree ensembles for estimating the causal effect of a continuous
  treatment of exposure which (i) does not require a priori parametric
  specification of the influence of control variables, and (ii) allows
  for identification of effect modification by pre-specified
  moderators.  The main parametric assumption we make is that the
  effect of the exposure on the outcome is linear, with the steepness
  of this relationship determined by a nonparametric function of the
  moderators, and we provide heuristics to diagnose the validity of
  this assumption.  We apply our methods to revisit a 2001 study of
  how abortion rates affect incidence of crime.
\end{abstract}

\vfill


\newpage
\spacingset{1.45} 


\section{Introduction}
\label{sec:intro}

\subsection{The abortion-crime hypothesis}

In the early 2000s, \cite{Levitt} published a paper claiming that the
legalization of abortion in the United States in the 1970s played a
decisive role in the dramatic reduction of crime incidence there in
the 1980s and 1990s.  The authors' main hypothesis is that the
legalization of abortion nationwide with the Supreme Court's 1973
decision in \textit{Roe v. Wade} led to a reduction in the number of
unwanted births, and that the children from such unwanted births are
at an elevated risk for criminal involvement beginning at the time
they reach adolescence.  Therefore, once abortion became legalized and
gradually more common, children aging into adolescence and adulthood
in the 1980s and 1990s contained fewer individuals with an greater
propensity to commit crimes.  The authors present empirical evidence
for the negative relationship between the lagged abortion rate and the
incidence of murder, violent crime, and property crime.

This finding has drawn considerable attention and controversy since
its initial publication.  It was featured in a highly popular book
\citep{freakonomics} and has been the recipient of much scrutiny from
the broader academic community; see the references in the
retrospective study by \cite{LevittNew} for a comprehensive list of
replies.

Notably, there have been numerous papers which question the original
paper's modeling assumptions in controlling for confounding factors
that underpin their results.  To estimate the causal effect of
abortion on crime, \cite{Levitt} account for social, economic, and
policy variables which may explain variation in both abortion and
crime rates.  Each of these variables is included as a term in a
linear model, and then the effect of abortion on crime is estimated
through ordinary least squares and the use of robust standard errors.
One main criticism of this approach is that the usage of such a linear
model may omit important higher-order trends in the control variables
affecting crime, leading to an inconsistent estimate of the causal
effect.  Several reanalyses of \cite{Levitt}'s data use more complex
models which introduce an expanded set of control terms, for instance
by interacting state and year dummy variables to allow for
state-specific temporal trends and interacting the year with the
original socioeconomic control variables to allow for temporal trends
in the controls, and claim that doing so explains away the original
finding \citep{FooteGoetz,belloni2014inference,RIC}.

Relatedly, \cite{woody2020confounder} performed a sensitivity analysis
using the data from the 2001 study, starting with the expanded linear
model specification from \cite{RIC} and considering inference for the
treatment effect under nested subsets of control terms.  They found
that the effect of abortion on crime is found to be null only with the
inclusion of quadratic temporal trends in the covariates and for each
state.  The intuition for this is clear: the progressive introduction
of terms to the linear model gradually reduces precision of the
estimated treatment effect, eventually leading to uncertainty
intervals (either confidence intervals and Bayesian posterior credible
intervals) which encompass 0.  Generally, one must be conservative in
model specification so as to meet the condition of ignorability, or
exogeneity in the econometrics literature.  However, this finding by
\cite{woody2020confounder} raises the possibility that the original
finding was valid, and that subsequent reanalyses simply added a
sufficient number of terms into the linear model to the point that any
significant effect was washed away.  In fact, a recent retrospective
study by \citeauthor{Levitt} using data updated with 17 years of data
following the 2001 publication found that their original prediction,
that crime would continue to fall as a lagged effect of increased
abortion rates, did indeed come to pass \citep{LevittNew}.


This is one particular instance in the recent reproducibility crisis,
whereby original published findings do not appear to hold up once
subjected to independent reanalysis.  This phenomenon has especially
afflicted the social sciences.  \cite{bryan2019} argue that the
flexibility exercised in the design and analysis stages by researchers
attempting to replicate a published result, so-called ``replicator
degrees of freedom,'' tends to result in a bias toward false-negative
replication results.  In this example, the major degree of freedom is
the model specification used for estimating the impact of abortion on
crime, especially how the control variables influence the outcome.

To address these concerns, we present our own reanalysis of the data
from \cite{Levitt} using a novel Bayesian semiparametric regression
model which can identify the causal effect of a continuous exposure
variable while obviating the need for explicit parametric
specification for the role of control variables.  This model is a
direct extension of the model from \cite*{bcf} for binary treatments,
though here it is adapted for continuous treatments.  Our model allows
for possible nonlinear and interactive effects of control variables in
the model for the outcome by incorporating them into a nonparametric
control function.  This feature of our model helps reduce researcher
degrees of freedom in analyses of primary studies and reduce
replicator degrees of freedom in replication studies, thereby
improving the robustness and reproducibility of published~results.

Our model presents one further contribution by accommodating
hetereogenity in the exposure effect by pre-exposure moderators.
Estimation of heterogeneous treatment effects has become a major focus
in the causal inference literature.  Detecting possible unanticipated
treatment modification can generate novel hypotheses, especially
concerning the mechanism of the treatment effect, to be tested in
subsequent studies \citep{bryan2019,tipton2020}.  The main parametric
assumption of the model is that the effect of the exposure on the
outcome is linear; we model heterogeneous treatment effects by
allowing the slope of the linear treatment effect to rise or fall as a
function of the moderators.  Keeping with the spirit of reducing
researcher degrees of freedom, we also specify a nonparametric form
for this effect moderation function.  To prevent the spurious
identification of effect moderation, we heavily regularize this
function so as to give conservative estimates of
hetereogenity. 

We apply our model to the data from \cite{Levitt}, focusing on the
causal effect of abortion on the murder rate in the 48 contiguous
United States from 1985 to 1997 (we also analyze abortion's effect on
violent crime and property crime using the same model, and present
these results in the supplement).  We provide evidence for a negative
causal relationship between abortion and murder rate, consistent with
the conclusions of \cite{Levitt}; this finding also holds for violent
crime and property crime.  Importantly, we demonstrate that the
inclusion of heterogeneous treatment effects does not significantly
alter inference on the average treatment effect (ATE), which is the
most common estimand in the literature, including the line of work on
the abortion-crime hypothesis started by \cite{Levitt}.  In other
words, we can gain greater insight into the effect of abortion on
crime without biasing inference on the ATE or formulating a second
model that includes interactions between the exposure and the
moderators, the fitting of which can itself induce bias.  We also
demonstrate that there is important between-state hetereogenity in the
effect of abortion on crime, and that more generous distribution of
welfare benefits to low-income families possibly diminishes this
effect.  This provides an opening for possible further research into
the dynamic between abortion and crime.

There have been several
other papers which have utilized causal inference to estimate the
effect of policies on the incidence of crime, most of which analyze
the relationship between gun regulation legislation and the rate of
gun violence.  For instance, \cite{Hasegawa2019} study the impact that
a repeal of a permit-to-purchase law in Missouri had in increasing gun
homicides; \cite{small2019after} evaluate the impact of police removal
of guns from domestic abusers; and \cite{schell2018evaluating} review
the impact of several gun access and use laws in lowering gun deaths.

The rest of the paper proceeds as follows.  For the remainder of this
section, we review the literature for the most relevant related
methodolgy.  In Section~\ref{sec:methods} we present our model and the
identification assumptions necessary to estimate the causal effect of
continuous treatments and exposures.  In Section
\ref{sec:analys-donah-levitt} we present our main application,
studying the impact of abortion on the murder rate as first done by
\cite{Levitt}.  A key parametric assumption embedded in our model is
that the treatment effect is linear, conditional on the value of
moderator covariates, and we conclude our application by presenting a
model check diagnostic to assess the validity of this assumption.  In
Section~\ref{sec:experimental-results} we explore this model check
diagnostic at length, presenting a case where it can detect the
existence of nonlinearity in the treatment effect even with the
presence of confounding.  We conclude with a brief discussion in
Section~\ref{sec:discussion}.

\subsection{Background}

\subsubsection*{Heterogeneous treatment effects}

Our method is developed with the motivation to detect and communicate
heterogenous treatment effects.  While much of classical causal
inference focuses on estimating average treatment effects, there is a
substantial and growing interest in estimating how treatment effects
vary according to unit-specific characteristics.  Estimation of
heterogeneous treatment effects has become a major focus in a diverse
array of research areas, including subgroup discovery in clinical
trials \citep{sivaganesan2017subgroup}, education and other public
policy fields \citep{yeagernature,finucane2015works}, and targeted
advertising \citep{gaines2011experimental}.  Many modeling
advancements have been made recently for estimating treatment effect
variation, including outcome modeling using Bayesian additive
regression trees \citep{hill2011bayesian,bcf,starling2019targeted},
using randomization-based inference \citep{ding2019decomposing}, and
matching techniques \citep{lee2018discovering,hsu2013effect}.
\cite{carvalho2019assessing} present the results of a recent workshop
in which eight invited teams analyzed a common dataset to detect
treatment effect variation.

\subsubsection*{Bayesian tree models}

The model we present makes heavy use of Bayesian tree ensembles, as
first introduced for regression with the method of Bayesian additive
regression trees \cite*[BART;][]{chipman2010}.  There exists a rich
literature of Bayesian tree models, which a cover wide range of topics
including survival analysis
\citep{sparapani2016nonparametric,sparapani2020nonparametric}; BART
models which adapt to smoothness
\citep{linero2018bayesian2,starling2020bart}; heteroskedastic outcomes
\citep{pratola2016efficient}; variable selection
\citep{linero2018bayesian,liu2018variable} categorical and count data
\citep{murray2017log}; gamma and inverse-gamma regression
\citep{linero2020semiparametric}; and, as mentioned, causal inference
\citep{bcf,hill2011bayesian,starling2019targeted}.  See
\cite*{hill2020bayesian} for a recent comprehensive review of modeling
developments using BART.

\subsubsection*{Posterior summarization}

We estimate heterogeneous treatment effects through the use of a
nonparametric function.  In order to directly and qualitatively
characterize this heterogeneity, we use the method of posterior
summarization \citep{woody2019model}, whereby complex
large-dimensional functions are interpreted by projecting them down
onto simpler structures.  This idea originates from the idea of
separating model fitting and interpretation, as first suggested by
\cite{maceachern2001decision} and further expanded upon by \cite{DSS},
rather than allowing the eventual goal of interpretation to drive
formulation of the model.  Similar ideas as applied to causal
inference can be found in the work \cite*{woody2020confounder},
\cite*{sivaganesan2017subgroup}, and \cite{carnegie2019examining}.


\section{Methods}
\label{sec:methods}

\subsection{Identification conditions}
\label{sec:ident-cond}

We use the potential outcomes framework \citep{IR2015} to analyze the
causal effect of a continuous exposure or treatment
$Z \in \mathcal{Z} \subseteq \mathbb{R}$ on some outcome of interest
$Y$, conditional on a sufficient set of possible confounders $X$.
That is, we are interested in comparing potential outcomes $Y(z)$,
which is the outcome when the exposure is set to $Z = z$.  We make
three main identifying assumptions:
\begin{enumerate}[(i)]
\item \textit{Consistency}: $Z = z$ implies $Y = Y(z)$
  \citep{rubin1978bayesian}
\item \textit{Weak unconfoundedness}: $Y(z) \ind Z \mid X$ for all
  $z \in \mathcal{Z}$ \citep{Imbens2000}.
\item \textit{Positivity}: $ \pi(z \mid x) > 0$ for all
  $z \in \mathcal{Z}$ where $\pi(z \mid x)$ is the conditional density
  for the treatment given the covariates, sometimes called the
  generalized propensity score \citep{Imbens2000, Hirano2004}.
\end{enumerate}

\subsection{Model definition}
\label{sec:model-definition}

In the case of a binary exposure or treatment, i.e. $Z \in \{0, 1\}$,
the most common scenario considered in the causal inference
literature, causal estimands are nearly always comparisons of
$Y(Z = 1)$ against $Y(Z=0)$.  In contrast, for continuous exposures
there is a more rich set of possible casual estimands.  One such
estimand is the finite difference average treatment effect (ATE),
\begin{align}
  \label{eq:ATE}
  \ATE_{z', z}(x) = \E[Y(z') - Y(z)] 
\end{align}
for any two levels of treatment of interest $z', z \in \mathcal{Z}$,
where the expectation is taken over the observed units.  Relatedly,
the dose-reponse function is the expectation of the potential outcome
as a function of the exposure
\begin{align*}
  \phi(z) = \E[Y(z)],
\end{align*}
from which \eqref{eq:ATE} may be estimated comparing any two levels of
treatment.

Recently there has been much interest in the causal inference
literature to estimate treatment effect heterogeneity, where the
treatment effect is moderated by some unit-specific covariates $x$.
The relevant causal estimand in this case is the finite difference
conditional average treatment effect (CATE), defined by
\begin{align}
  \label{eq:CATE}
  \CATE_{z', z}(x) = \E[Y(z') - Y(z) \mid X = x]. 
\end{align}

In this paper, estimation of treatment effect variation via the CATE
function is a primary motivation.  To do so, we propose the following
semiparametric regression model:
\begin{align}
  \label{eq:model}
  y &= \mu(x_\C) + \tau(x_\M) \cdot z + \varepsilon,
      \quad \varepsilon \sim \mathcal{N}(0, \sigma^2),
\end{align}
where $\mu(\cdot)$ and $\tau(\cdot)$ are both nonparametric functions
each represented as a sum of Bayesian regression trees, as first
introduced by \cite*{chipman2010}.  This model is semiparametric in
the sense that the two functions $\mu(\cdot)$ and $\tau(\cdot)$ are
nonparametric, but we make the parametric assumption that, conditional
on covariates $x$, the relationship between $y$ and $z$ is linear.
The steepness of this linear relationship is moderated by the
$\M$-subset of covariates in $x$ as determined by the function
$\tau(\cdot)$.  Hence, $\tau(\cdot)$ is called the \emph{exposure
  moderating function}, and the covariates $x_\M$ are called
moderators.  Meanwhile, the baseline expected outcome for a fixed
level of the exposure $z$ is given by $\mu(\cdot)$, which we call the
\emph{control function}, and the covariates $x_\C$ are called the
control variables.  The control variables need to be specified so that
the condition of weak unconfoundedness holds.  In practice, $x_\C$ and
$x_\M$ may be specified to be identical, overlapping, or disjoint sets
of covariates.

It is well-known that estimation of causal effects necessitates the
conditioning on a sufficient set of control variables such that the
assumption unconfoundedness is met.  However, even if this assumption
is met, it is usually necessary to specify the role of these controls
in affecting the outcome, usually codified through the specification
of parametric assumptions in the outcome model.  This is closely
related to the topic of confounder selection
\citep{zigler2014uncertainty,wilson2014confounder,wang2015,RIC}.  Our
model obviates the need for such strict specification through the use
of a nonparametric control function.

Using the model in \eqref{eq:model}, the CATE has the explicit form
\begin{align}
  \label{eq:CATE-model}
  \CATE_{z + \Delta z,z}(x) = \tau(x_\M) \cdot \Delta z,
\end{align}
that is, conditional on a vector of moderators $x_\M$ the CATE is
linear across all values of $z$, with the slope given by $\tau(x_\M)$.
Because of the linearity assumption, the average treatment effect is
given by
\begin{align*}
  \text{ATE} \equiv \bar \tau = N^{-1}\sum_{i=1}^{N} \tau(x_{i, \M}),
\end{align*}
and when we refer to the treatment effect for a particular unit, we
mean $\tau(x_{i,\M})$, the value of the moderating function evaluated
at the unit's specific vector of moderators.

In this paper we are directly interested in estimating treatment
effect variation, rather than estimating the dose-response function.
For this we refer the reader to the rich existing literature
\cite[e.g.][]{Hirano2004,kennedy2017non,incremental,moodie2012estimation,wu2020matching}.

\paragraph{Prior specification}

For the regression model in \eqref{eq:model} we use a prior mostly
aligned with that from \cite*{bcf}, whose model we also closely
emulate.  Importantly, we regularize the sum of trees for the
$\tau(\cdot)$ function more heavily, i.e. to use much more shallower
trees, so that the model does not detect spurious exposure effect
heterogeneity.  The biggest difference between our model and that of
\cite{bcf} is that our model is meant for use with continuous, rather
than binary, exposure, meaning that we must carefully specify the
prior for the scale of $\tau(\cdot)$ which controls the slope of the
treatment effect.  For this, we use a half-Gaussian prior, though this
choice can be also be made empirically.  Finally, for the observation
error variance, we use Jeffreys' prior.


\section{Application: revisiting the impact of abortion on crime}
\label{sec:analys-donah-levitt}

\subsection{The data}

For this analysis, we consider the causal effect of abortion on
homicide rates across the United~States.  \cite{Levitt} and subsequent
authors also inspect the impact of abortion on the rates of property
crime and violent crime.  Here we restrict our attention to homicide
because our analysis shows that it is most heavily affected by
abortion out of these three outcomes, though we replicate our analysis
for violent crime and property crime in the supplement.

The outcome considered is $y_{st}$, the log per-capita murder rate
observed for state $s$ during year $t$, and the exposure of interest
is $z_{st}$, the ``effective abortion rate'' abortion rate (quantified
by number of abortions per live birth).  The effective abortion rate
was defined by \cite{Levitt}, and time-lags and weights the abortion
rate from previous years according to the age distribution within the
cohort of criminal offenders for the current year. That is, if 30\% of
homicide offenders in year $t$ are of age 18 and 70\% are of age 19,
then the effective abortion rate for year $t$ is
$0.3 \times \text{abortion-rate}_{t-18} + 0.7 \times
\text{abortion-rate}_{t-19}$.

Our dataset consists of recorded values of the outcome and exposure
for the 48 contiguous states in the United States (i.e., we exclude
Alaska, Hawaii, and the District of Columbia from consideration) for
the years 1985 to 1997 (inclusive).  This yields a total of $N = 624$
observations.  For notational simplicity, we index the observations by
$i=1,\ldots,N$, and then define a vector of state dummy variables
$s_i$ (e.g. $s_{ik} = 1$ if observation $i$ belongs to state $k$,
$k=1,\ldots47$, and $s_{ik'}=0$ for $k' \neq k$) and denote the year
by $t_i=1985,\ldots,1997$, which we use as a scalar numeric value.
Furthermore, for each observation we have a vector of eight
covariates, which we collectively denote by $x_i$.  We describe each
of these covariates in Table~\ref{tab:covariate-description}.

\begin{singlespacing}
    \begin{table}[tb] \footnotesize
      \centering
      \begin{tabular}[ht]{  l | p{0.3\textwidth} | l | l} \toprule
        Covariate   & Description & Used as control? & Used as moderator? \\ \midrule
        \texttt{state} & categorical variable for state (contiguous US states; 48 levels) & Yes & Yes \\ \midrule
        \texttt{year} & numeric value for year (1985--1997, inclusive) & Yes & Yes \\ \midrule \midrule
        \texttt{police} & log-police employment per capita & Yes & No \\ \midrule
        \texttt{prison} & log-prisoner population per capita & Yes & No \\ \midrule
        \texttt{gunlaw} & indicator variable for presence of concealed weapons law & Yes & No \\ \midrule
        \texttt{unemployment} & state unemployment rate & Yes & Yes \\ \midrule
        \texttt{income} & state log-income per capita & Yes & Yes \\ \midrule
        \texttt{poverty} & state poverty rate & Yes & Yes \\ \midrule
        \texttt{afdc15} & generosity to Aid to Families with Dependent Children (AFDC), lagged by 15 years & Yes & Yes \\ \midrule
        \texttt{beer} & beer consumption per capita & Yes & Yes \normalsize \\ 
        \bottomrule
      \end{tabular}
      \caption{Descriptions of covariates used as controls and
        moderators for analysis of Donahue and Levitt data.  Control
        variables are used in the control function $\mu(\cdot)$, and
        moderator variables are used in the exposure moderating
        function $\tau(\cdot)$ as denoted by $x_\C$ and $x_\M$,
        respectively, in the model equation \eqref{eq:levitt-model}.
      }
      \label{tab:covariate-description}
    \end{table}
\end{singlespacing}


\subsection{The model}
\label{sec:model}

These data come from an observational study rather than a randomized
experiment, and so it is important to consider the effect of
confounding factors. The eight covariates contained in
Table~\ref{tab:covariate-description} capture socioeconomic and policy
factors which could possibly explain variation in both homicide and
abortion, and so are potential confounders.  In addition, there could
be state-level effects or temporal trends in homicide which confound
the relationship between abortion and homicide.  
Therefore, following \cite{Levitt} and subsequent analyses of
these data, we use each of these covariates, as well as the state
dummies $s$ and the year $t$, as control variables.

In contrast to previous analyses of these data, we allow for the
possibility that the effect of abortion on crime varies across states,
over time, and through a subset of the covariates.  To do so, we
selectively include certain covariates into the exposure moderating
function.  We exclude variables \texttt{police}, \texttt{prison}, and
\texttt{gunlaw} from the set of moderators because seems intuitive
that these are far less likely to modify the impact of abortion on
crime.  Therefore, the set of the moderators in our model includes
\texttt{afdc15}, \texttt{income}, \texttt{poverty},
\texttt{unemployment}, and \texttt{beer}, as well as the state dummies
$s$ and year $t$.  However, our results do not change dramatically
when we include all the recorded covariates as moderators.

Therefore the model we use for our analysis is fully described by
\begin{align}\label{eq:levitt-model}
  y = \mu(x_{\C}, s, t) + \tau(x_\M, s, t) \cdot z + \varepsilon,
  \quad \varepsilon \sim \N(0, \sigma^2) 
\end{align}
where $\C$ and $\M$ are the sets of covariates used as controls and
moderators, respectively, as delineated in
Table~\ref{tab:covariate-description}.

This model shares one core feature to those used in previous analyses:
we assume that, conditional on observed covariates, the causal
relationship between abortion and homicide is linear.  However, our
model makes two significant departures from those specified by
previous authors.  First, instead of making the strict parametric
assumption of a linear relationship between the outcome and each
control variable through the use of a linear model, we allow for more
complex trends (nonlinearities and interactions) among these control
variables by including them in the nonparametric control function
$\mu(\cdot)$.  Second, we allow for heterogeneity in the causal effect
through the use of the moderating function $\tau(\cdot)$.  This
captures the possibility that the causal effect of abortion on crime
is stronger in some states than in others, that it is time-dependent,
or that it is amplified or mitigated by some of them moderator
covariates.  Again, the model is agnostic regarding the particular
form of effect modification by these moderators. 

\subsection{Results}

In order to perform posterior inference on the model in
Eq.~\eqref{eq:levitt-model} conditional on observed data, we used a
Gibbs sampler to generate Monte Carlo draws from the posterior using
the \texttt{multibart} package \citep{multibart} in the R programming
language \citep{R}.  To ensure a sufficient degree mixing, we ran 100
Markov chains in parallel, taking 500 samples from each chain after
discarding the first 25,000 samples as a burn-in.  This gave a total
of 50,000 posterior samples.

\subsubsection{Average treatment effect}

\begin{figure}[t!]
  \centering
  \includegraphics[width=0.9\textwidth]{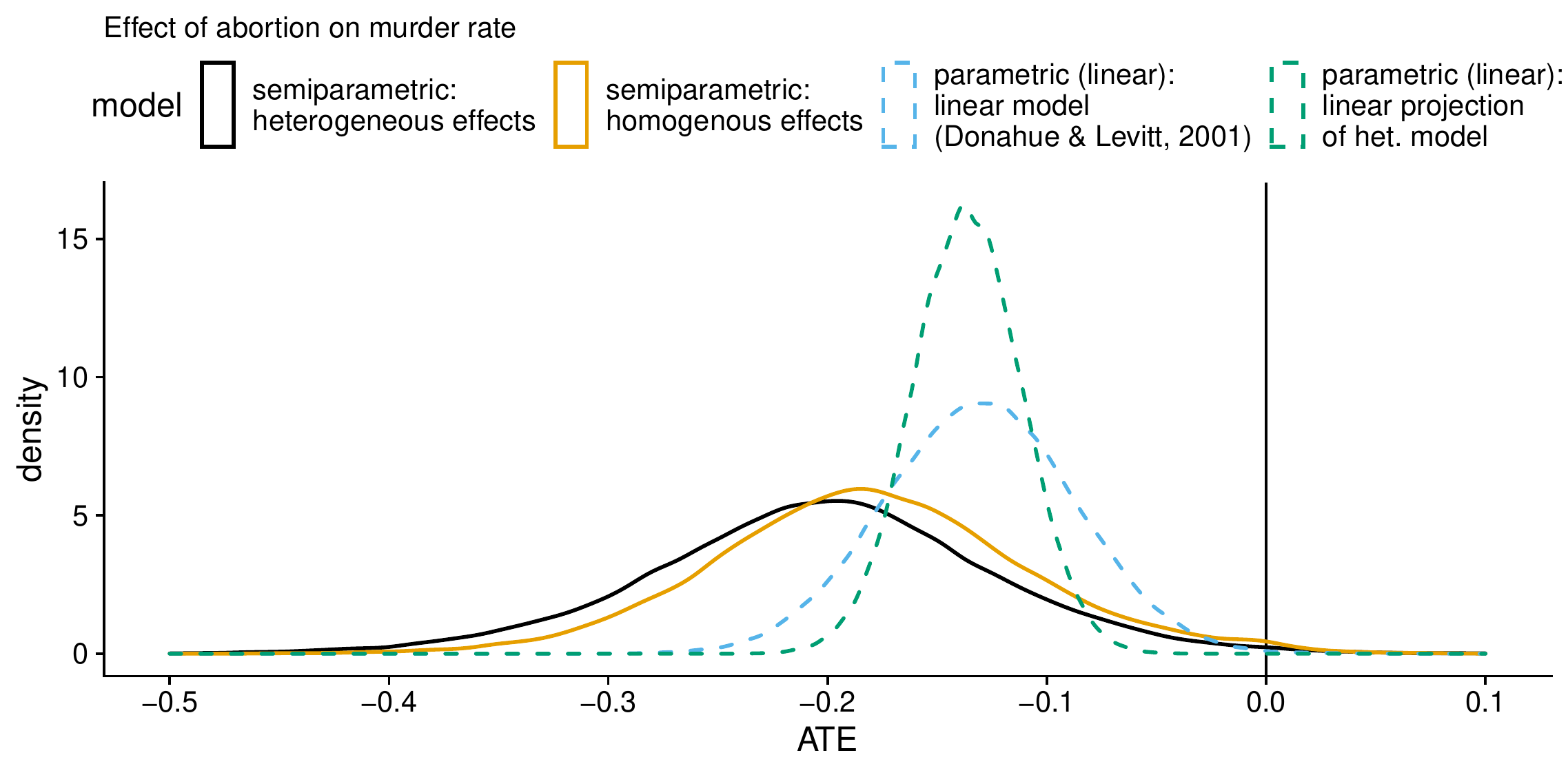}
  \caption{\label{fig:levitt-ate} Estimates of the average treatment
    effect (ATE) for the analysis of the Donahue and Levitt murder
    data.  We compare the posterior for the ATE from the
    semiparametric models which allow for heterogeneous and
    homogeneous treatment effects.  We also show the confidence
    interval for the treatment effect using the linear model
    specification from the original paper by \cite{Levitt}, and the
    projected posterior for the ATE found by projecting the ATE
    posterior from the heterogeneous effects model onto the same
    linear structure.  }
\end{figure}

First, we perform inference on the average treatment effect (ATE),
which is the sample average of the individual treatment effects,
$\bar \tau \equiv N^{-1}\sum_{i=1}^{N}\tau(x_{i,\M})$.
Figure~\ref{fig:levitt-ate} shows the posterior for the ATE under our
heterogeneous effects model in Eq.~\eqref{eq:levitt-model}.  For
comparison, we also show the posterior for the ATE under a homogeneous
effects model, when there is a nonparametric control function
$\mu(\cdot)$ but no treatment moderation.  This is equivalent to the
use of the model in Eq.~\eqref{eq:levitt-model} under the special case
where $\tau(\cdot) \equiv 1$. The posterior mean and 95\% credible
interval for the ATE under the heterogeneous effects is --0.203
(--0.365, --0.047), and under the homogenous effects model is --0.181
(--0.322, --0.029).  Thus, the posterior mean under the homogenous
effects model is only about 10\% smaller in magnitude compared to that
from the heterogenous effects model, with a credible band about 9\%
wider; neither credible interval straddles 0.  This demonstrates that
accounting for treatment effect moderation does not significantly
alter inference on the average treatment effect for the analysis of
these data.

Next, we compare the ATE estimate from our semiparametric model
against estimates arising from the use of a fully linear model
specification.  Figure~\ref{fig:levitt-ate} also shows the posterior
for the ATE when using the same linear model specification from
\cite{Levitt} with a flat prior on all coefficients (in their paper,
\citeauthor{Levitt} use ordinary least squares and report robust
standard errors).  This linear model specification assumes a linear
relationship between the outcome and the eight control covariates in
Table~\ref{tab:covariate-description}, and also includes dummy
variables for both state and year, while not accommodating
heterogeneous treatment effects.  The resulting posterior from the use
of a flat prior and a linear structure is shown by the blue dashed
line in Figure~\ref{fig:levitt-ate}.  This model gives a posterior for
the ATE which has a point estimate notably smaller in magnitude and is
more peaked compared to that from the heterogeneous effects model,
though the two posteriors still show some overlap.

One problem that arises is that this ATE estimate from refitting the
model with a linear specification constitutes a second use of the
outcome data, and so we therefore lose a strict Bayesian
interpretation of the resulting posterior.  To reconcile inference
between these two specifications, we fit a linear \textit{summary} of
the heterogeneous effects model as suggested by \cite{woody2019model,
  woody2020confounder}.  That is, we calculate the posterior for the
ATE when projecting the posterior of the heterogeneous effects model
down onto the linear structure used by \cite{Levitt}.  Algorithmic
details are contained in \cite{woody2020confounder}.  This projected
posterior is shown by the green dashed line in
Figure~\ref{fig:levitt-ate}.  We see that point estimates for the ATE
from the linear projection is mostly consistent with the point from
refitting the model with a flat prior.  The biggest difference is that
the projected posterior for the ATE is more peaked for the former than
for the latter.  There is a lesser degree of uncertainty in the ATE
estimate which stems from the use of the fitted heterogeneous effects
model, instead of using the higher-variance outcome data for a second
time to estimate the ATE.  Still, in either case the use of the linear
structure results in a large deviation in the estimated ATE away from
the ATE under the heterogeneous effects model, implying that this
linear structure is overly simplistic and does not adequately
recapitulate the role of the control variables.


\subsubsection{State-level effect heterogeneity}

Our results indicate that there is strong evidence pointing to a
negative effect of abortion on murder rates on average.  Still, this
conclusion on its own does not address potential heterogeneity in this
effect dependent on the state, year, and moderating covariates.  We
now shift our focus to addressing this question, in which we
characterize the posterior for the exposure effect moderating function
$\tau(\cdot)$.  It is important to note that the following analyses do
not refit the model in any way.  In other words, we do not ``double
dip'' by using the outcome a second time, but rather perform an
exploration of the posterior for $\tau(\cdot)$.  In this way, we
retain a strict Bayesian interpretation and valid posterior estimates
of inferential targets of interest.

\begin{figure}[t!]
  \centering
  \includegraphics[width=1\textwidth]{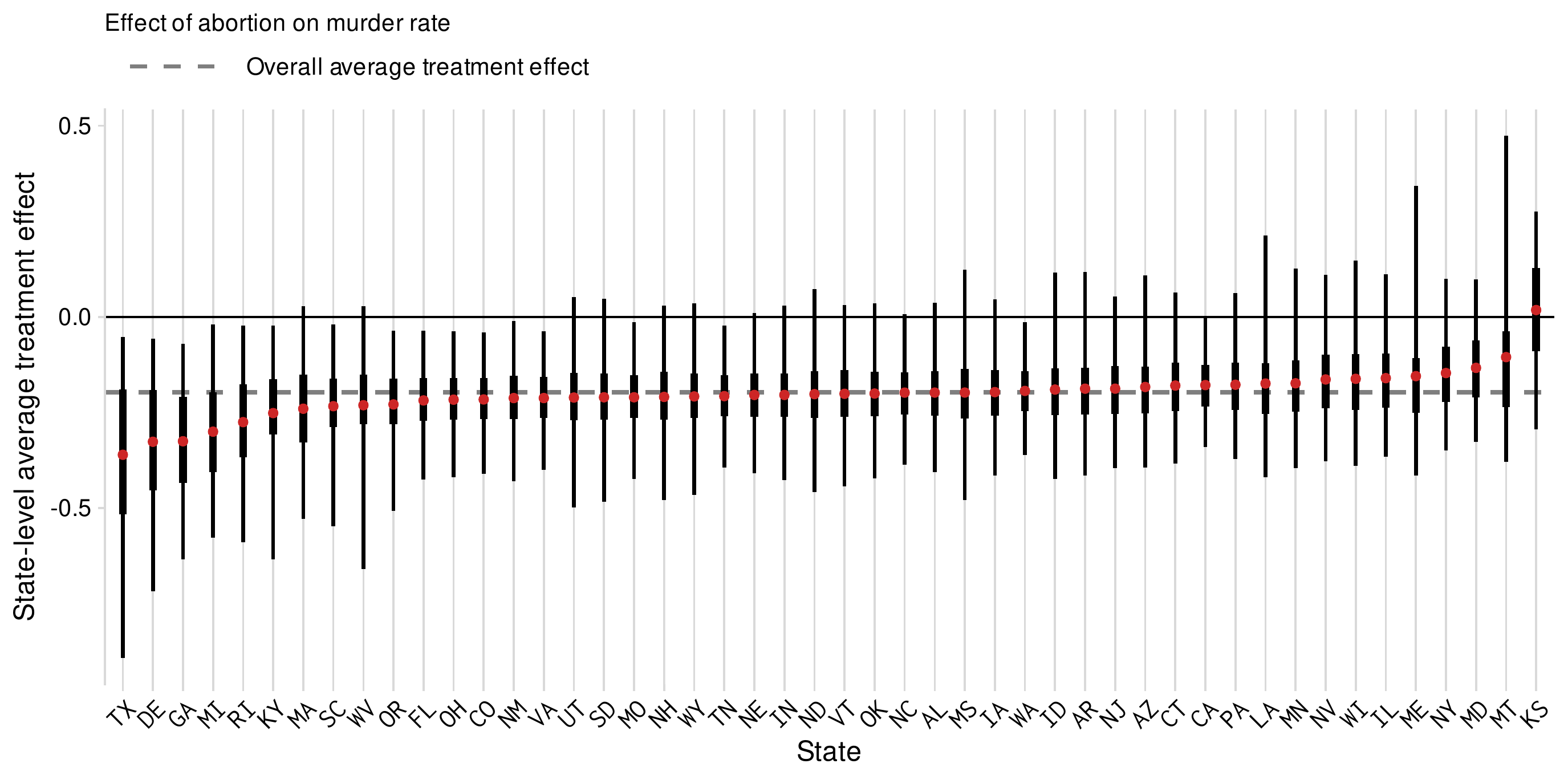}
  \caption{\label{fig:levitt-state} ATE estimate for each state, along
    with equal-tailed 50\% and 95\% credible intervals.  There is a
    considerable amount of variation in the effect of abortion on the
    murder rate among these 48 states.  For Texas, it appears to have
    a much stronger effect, whereas for Kansas it appears to have
    little to no effect. }
\end{figure}

First, we address between-state effect heterogeneity.
Figure~\ref{fig:levitt-state} shows the posterior for the state-level
ATE's, which we define to be the estimated causal effect for a given
state averaged across all the years included in the dataset.  That is,
the state-level ATE for state $k$ is
$\bar \tau_k = n_k^{-1}\sum_{i:s_i=k}^{}\tau(x_{i, \M}, s_i, t_i)$,
where $n_k = \sum_{i}1(s_i = k)$ is the number of observations for
state $k$.  From this we can see that abortion appears to reduce crime
at different levels among the states, in some cases quite
substantially. For some states, such as Texas, Delaware, and Georgia,
abortion has a greater effect on crime than average, while for others,
like Maryland and Montana, abortion has a lesser effect.  In fact, for
Kansas, the abortion rate seems to have little to no effect on the
murder rate.

\begin{figure}[t!]
  \centering
  \includegraphics[width=1\textwidth]{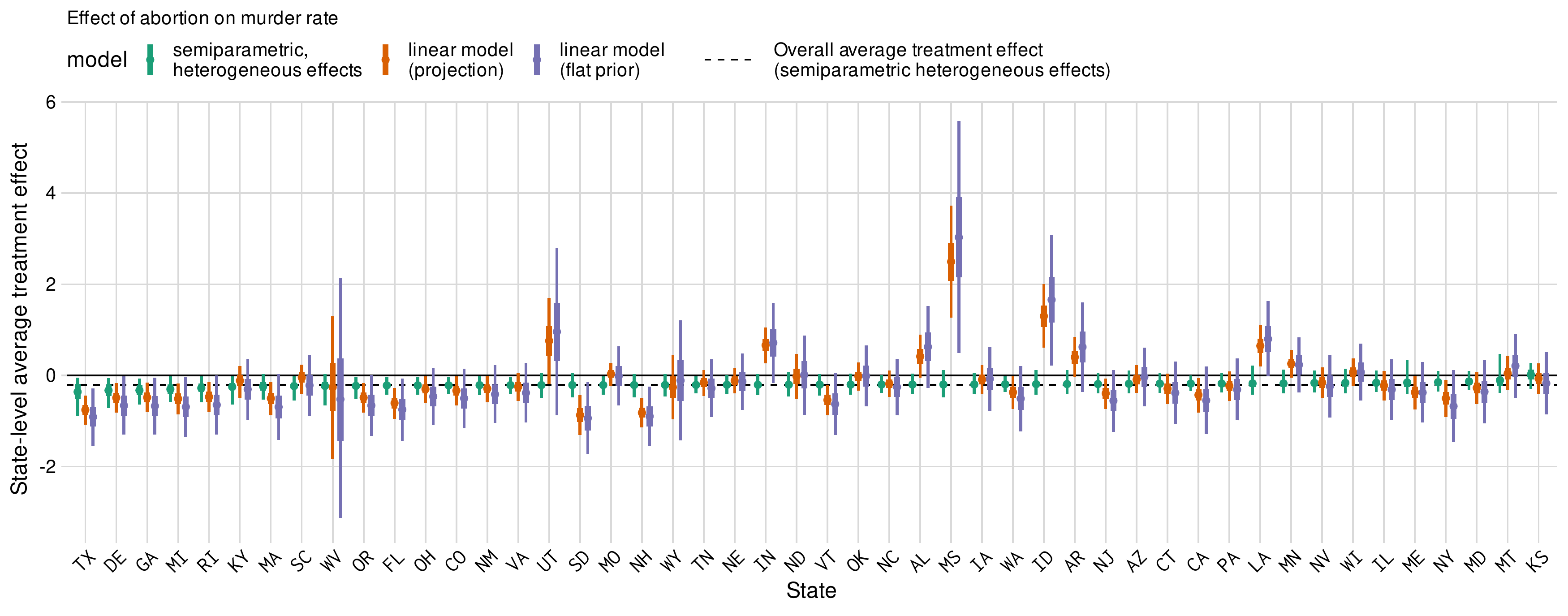}
  \caption{\label{fig:levitt-state-modelcomp} ATE estimate for each
    state, along with equal-tailed 50\% and 95\% credible intervals,
    comparing across three types of estimates: (i) heterogeneous
    effects model [green; same posteriors as in
    Figure~\ref{fig:levitt-state}], (ii) projecting the heterogeneous
    model onto linear structure with state-exposure interactions
    [orange], and (iii) refitting a linear model with state-exposure
    interactions with a flat prior [purple].  Both the linear
    projection and the refitted linear model show a high degree of
    variability compared to the heterogeneous effects model,
    reflecting the conservatism of our model in detecting effect
    modification.  }
\end{figure}

When we replicate our analyses for the other two crime outcomes
(property crime and violent crime, shown in the supplement) we even
see some suggestions that the exposure effect changes sign, becoming
positive when the overall ATE is negative.  This raises the
possibility of some unobserved state-level moderator which plays a
role in altering the effect of abortion on crime rates.  Still, the
state-level ATE's for the most part hem pretty closely to the overall
ATE, reflecting the inherent conservatism of our model which is
designed not to detect spurious heterogeneity.  

To underscore this feature of conservatism in estimating between-state
variance in the treatment effect, we show in
Figure~\ref{fig:levitt-state-modelcomp} two alternative ways to
estimate the state-level ATEs which both use a linear model structure
interacting the exposure with the state dummy variables.  The first
refits the model using the outcome data and a flat prior on all the
coefficients, while the second projects the posterior from the
heterogeneous effects model down onto this linear structure (as
described previously for estimating the overall ATE).  The estimates
from using the linear structure are very imprecise because of the
large dimensionality.  The projected posterior has slightly more
precise credible interval than those from the refitted model
posterior, but both are still rather large and have some questionably
high point estimates compared to the state-level ATE posteriors from
the heterogeneous effects model.  This suggests that our model gives
estimates of the overall ATE in line with estimates from simpler
models and with an appropriate level of uncertainty, while also
detecting a plausible level of between-state variation in the
treatment effect, especially when compared to an alternative linear
model specification.

\subsubsection{Detecting effect modification by covariates and over time}

We have seen that the magnitude of the causal effect of abortion on
murder rates appears to vary widely across states, but it is also
natural to ask how it is moderated by measured covariates and how it
varies across time.  The nonparametric form we specify for the effect
moderating function $\tau(x_{i, \M}, s_i, t_i)$ allows for arbitrary
nonlinearities and interactions between the moderating covariates,
state dummy variables, and year in altering the exposure effect.  This
feature is appealing because it allows us to be agnostic about the
specific contributions of these moderators to the heterogeneity in the
causal effect, thereby reducing researcher degrees of freedom which
may bias results in unanticipated ways.  However, an often cited issue
with such nonparametric functions is that they are seemingly opaque to
interpretation.

To address this issue and communicate significant trends within the
moderating function $\tau(\cdot)$ in a simple, interpretable manner,
we use the method of posterior summarization from
\cite{woody2019model} to investigate our estimate for $\tau(\cdot)$.
Posterior summarization is a method of explaining the predictive
features of high-dimensional, usually nonparametric, functions by
projecting them down onto lower-dimensional summary functions.
Because these summaries are functionals of the original
(nonparametric) function, their posteriors are implied by the
posterior for the original function.  Furthermore, because the data
are used only once, in computing the first-stage posterior, these
posteriors retain a valid Bayesian interpretation even across multiple
summaries.  In our case, we consider two separate yet related
summaries of the nonparametric effect moderating function
$\tau(\cdot)$.  First, we consider an additive summary, which reports
the average partial effect of each moderator in modifying the
treatment effect.  Second, we construct a tree summary of
$\tau(\cdot)$ which constructs disjoint subgroups of units with
contrasting conditional average treatment effects.

To begin, we first inspect the partial effect of time and each
moderating covariate in $\tau(\cdot)$, while still adjusting for
between-state variation in the exposure effect. This demonstrates how
the causal effect of abortion rates on the murder rate varies
according to these covariates.  We construct an additive summary,
whereby $\tau(\cdot)$ is projected down onto a lower-dimensional
(additive) summary function of the following form:
\begin{align}\label{eq:additive-function}
  \gamma(x_i, s_i, t_i)
  &= \alpha + \sum_{k=1}^{47} b_s \cdot 1(s_i = k) +
    \sum_{j=1}^{5} h_j(x_{ij}) + h_6(t_i).
\end{align}
Here $1(s_i = k)$ is an indicator function for observation $i$
belonging to state $k$, $k=1,\ldots,47$, and each $h_j(\cdot)$,
$j=1,\ldots,6$ is a smooth univariate function of one of the six
numerical treatment moderating variables (\texttt{afdc15},
\texttt{beer}, \texttt{income}, \texttt{poverty},
\texttt{unemployment}, and year) represented by a thin plate
regression spline with basis dimension 10 \citep{Wood2001}.  This
summary function communicates the average partial effect of each of
the six numerical moderators on $\tau(\cdot)$ through the additive
functions $h_j$, conditional on the state dummy variable.  Even though
the original function $\tau(\cdot)$ may contain interactions between
its inputs, this summary effectively averages over any interactions
which may be present.

The point estimate for the additive summary is the function which
minimizes the penalized squared difference between $\gamma(\cdot)$ and
$\tau(\cdot)$ evaluated at the all the observed values,
\begin{align}
  \hat \gamma(x) = \arg \min_{\gamma \in \Gamma}
  \sum_{i=1}^{N}[\hat \tau(x_i, s_i, t_i) - \gamma(x_i, s_i, t_i)]^2 + \sum_{j=1}^{6} \lambda_j J(h_j)
\end{align}
where $\Gamma$ is the set of functions of the form in
Eq.~\eqref{eq:additive-function} and
$J(h_j) = \int h_j''(u)\mathrm{d}u$ is a complexity penalty which
enforces smoothness in the fitted additive functions $h_j$,
$j=1,\ldots,6$.  We can also find a posterior for the summary by
projecting down Monte Carlo draws from the posterior of $\tau$ onto
this additive structure; algorithmic details are given by
\cite{woody2019model}.

\begin{figure}[t!]
  \centering
  \includegraphics[height=0.4\textheight]{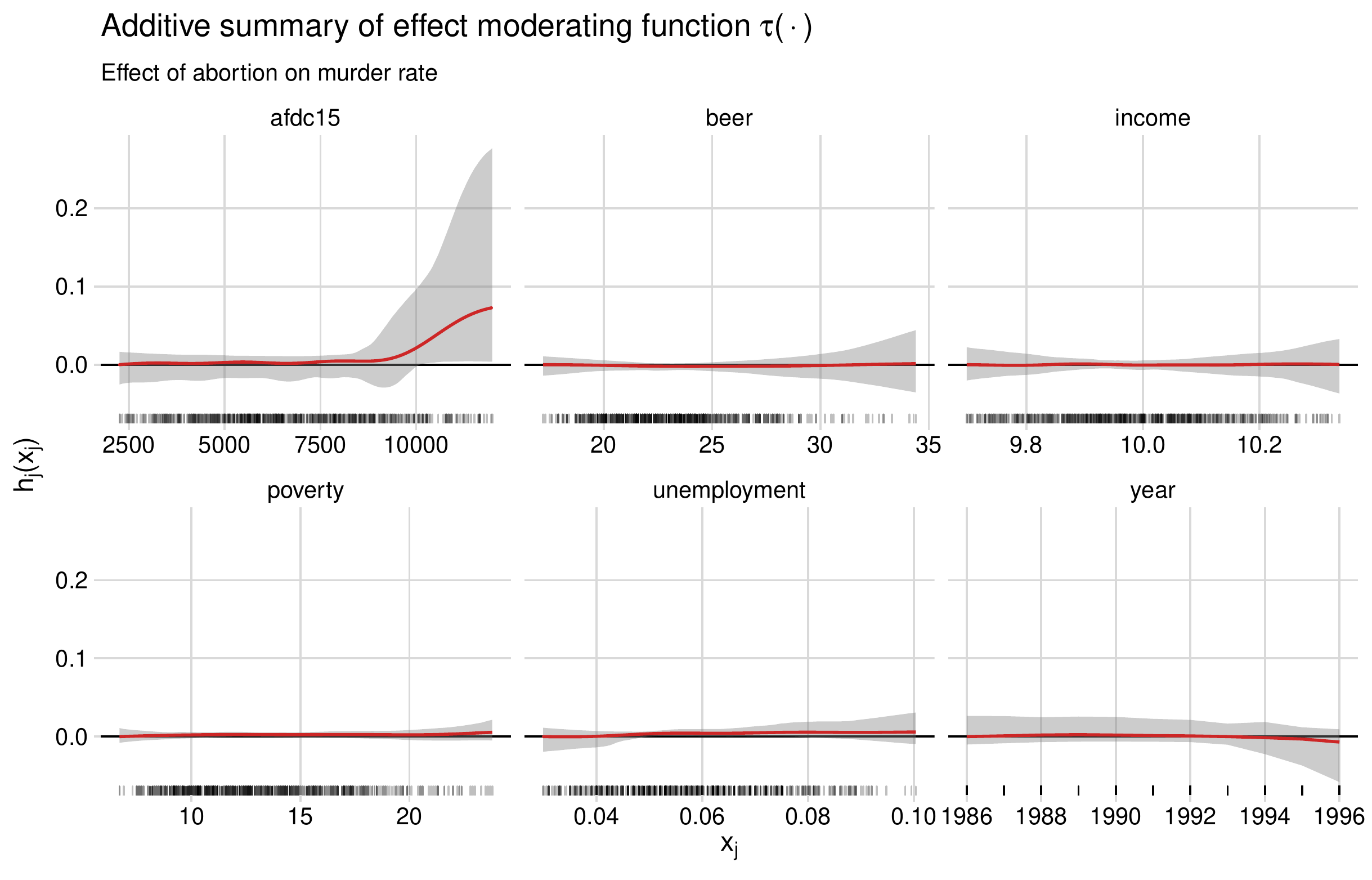} \\
  \includegraphics[width=0.35\textwidth]{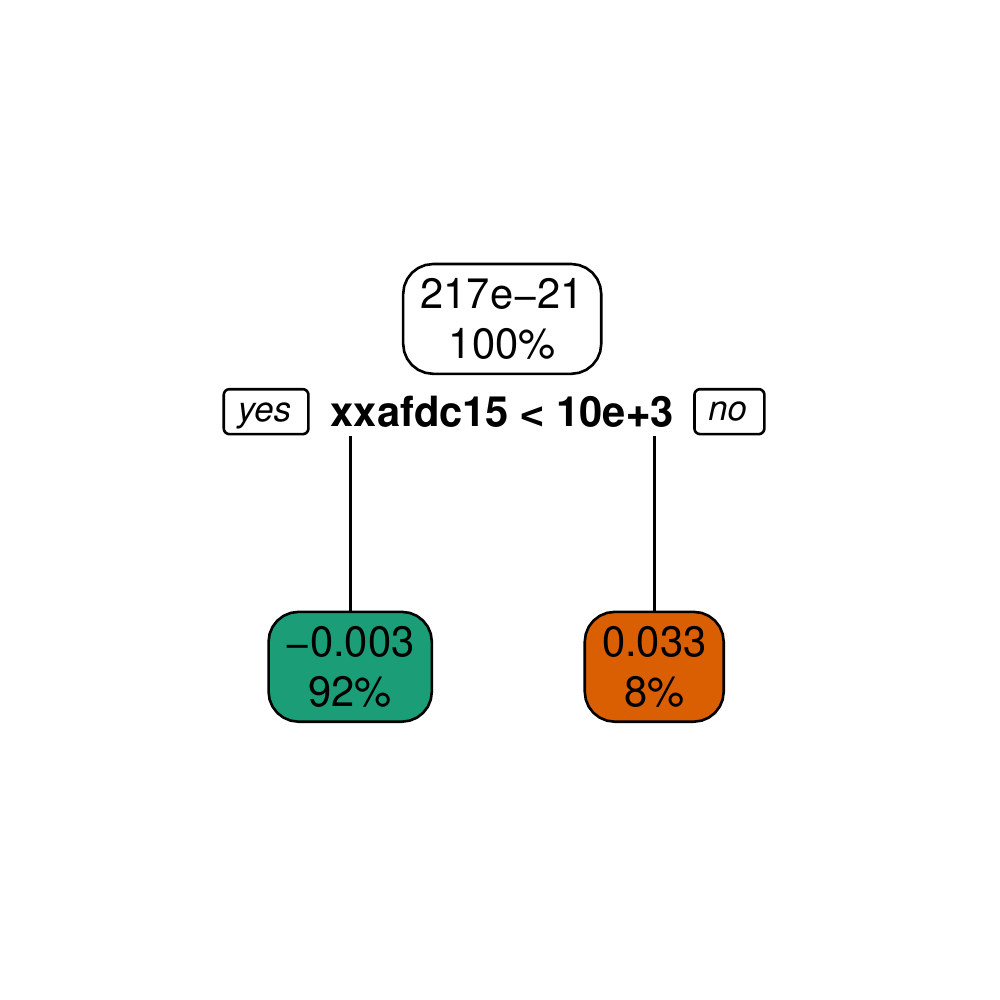}
  \includegraphics[width=0.64\textwidth]{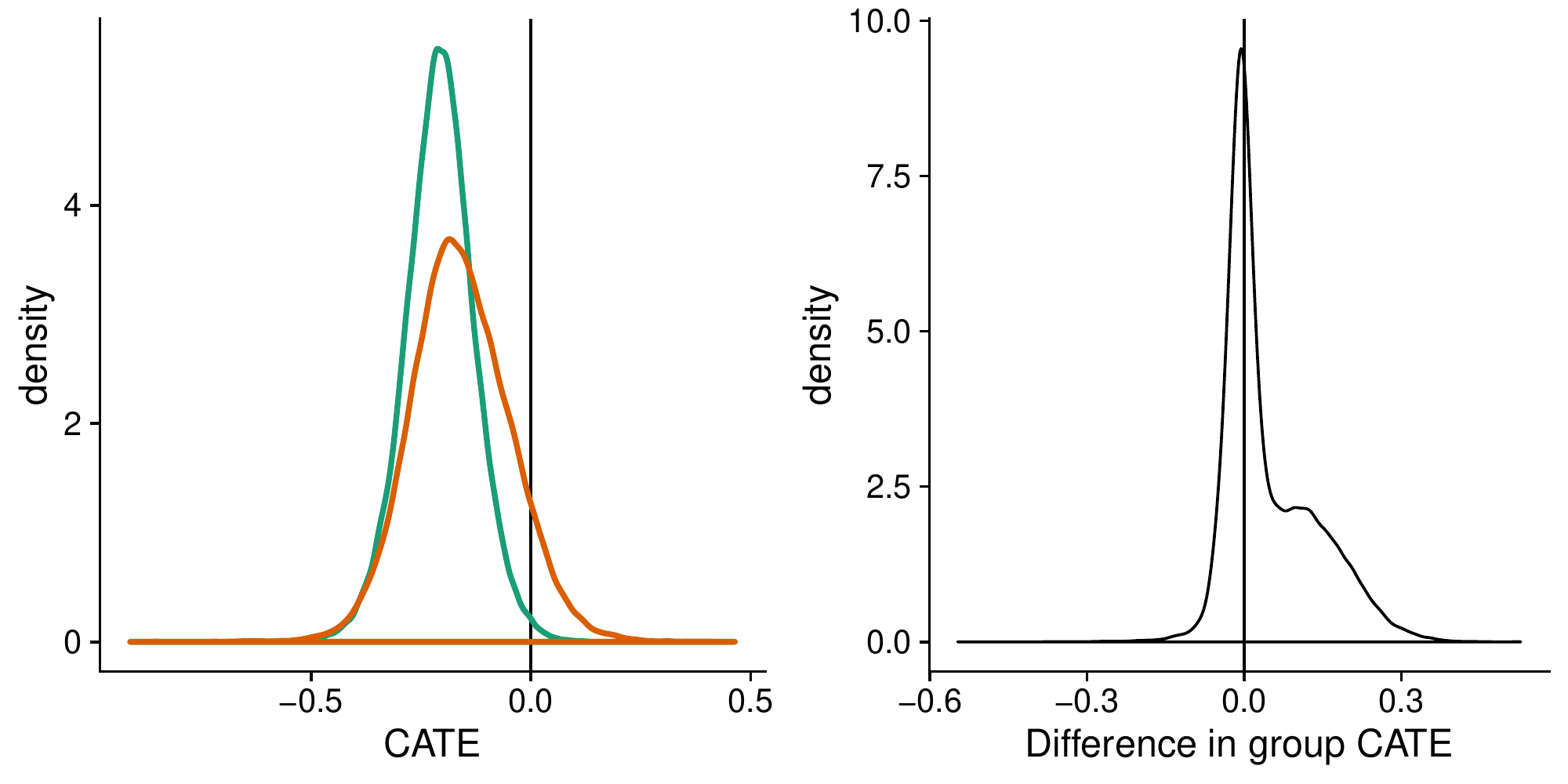}
  \caption{\label{fig:proj-tau-mu} Posterior summaries of the
    treatment modifying function $\tau(\cdot)$.  \textit{Top:}
    Additive summary for $\tau(\cdot)$, describing the average partial
    effect of each moderating variable on altering the treatment
    effect.  \textit{Bottom:} A regression tree trained to the
    $\hat \tau(x_i)$.  The tree contains a single split on
    \texttt{afdc15}, consistent with the additive summary.  This split
    results in two subgroups.  We show the posterior for the CATE for
    each of these two subgroups, and the posterior distribution for
    the difference in CATEs between them. }
\end{figure}

The fitted additive summary, along with 90\% credible intervals, is
shown in the top panel of Figure~\ref{fig:proj-tau-mu}.  The variable
with strongest influence on $\tau(\cdot)$ is \texttt{afdc15},
measuring the expenditures of the Aid to Families with Dependent
Children federal assistance program administered by the states, lagged
by fifteen years.  High levels of \texttt{afdc15} are associated with
an increase of the treatment effect by about 0.08 as compared to low
levels of \texttt{afdc15}.  This is quite a sizeable difference; all
else equal, this amounts to roughly a 40\% decrease of the treatment
effect in magnitude relative to the estimated average treatment
effect, though there is a wide interval around this shift.  However,
the other moderators have seemingly little influence on $\tau(\cdot)$,
nor does treatment effect tend to vary across time. This feature is
due to the inclusion of the state dummy variables in the additive
summary.  As shown in Figure~\ref{fig:levitt-state}, much of the
detected effect heterogeneity is likely due to variation between
states not explained by the moderators.  Once this variation is
accounted for in the additive summary, there is apparently little
remaining heterogeneity across years or the moderating covariates.

As a final step in our exploration of $\tau(\cdot)$, we attempt to
detect subgroups with contrasting conditional average treatment
effects (CATEs).  We do so by building a (single) regression tree
which regresses the individual treatment effects (ITEs) on the
moderating covariates $x_{i, \M}$.  To account for between-state
heterogeneity, we sweep out the state-level ATEs from the ITEs before
constructing the tree.

The bottom-left panel of Figure~\ref{fig:proj-tau-mu} shows the
resulting tree, which results in two subgroups determined by a single
split based on \texttt{afdc15}.  This replicates the finding in the
additive summary that the most significant moderator appears to be
\texttt{afdc15}.  This tree summary is also a functional of
$\tau(\cdot)$ because it projects the output of $\tau(\cdot)$ onto a
lower-dimensional function, in this case a summary function determined
by a series of binary decision nodes.  Therefore, the subgroups have
an implied posterior for their CATEs, found by dropping Monte Carlo
draws of the posterior for $\tau(\cdot)$ down the decision tree.  We
show the estimated posterior for these subgroup CATEs in bottom-middle
panel of Figure~\ref{fig:proj-tau-mu}, and also the posterior for the
difference between the CATEs in bottom-right panel.  The posterior for
the difference in CATEs contains a tall peak near zero, but also has a
wide right tail.  As we show in the supplement, the prior on this
difference has a sharp peak at zero.  Therefore, even though the
posterior still overlaps zero we can say that the right tail in this
posterior for the difference in CATEs indicates tentative evidence
supporting moderation by \texttt{afdc15}, in line with the results
from the additive summary.

All in all, we can conclude that there is strong evidence for a
negative causal effect of abortion on murder rates, and that there is
suggestive evidence that this effect is mitigated when a state has a
more generous welfare program for families with dependent children
which had little income.  This prospective effect modification by
\texttt{afdc15} lends insights into the mechanism of the causal
effect.  For instance, it could be the case that higher AFDC
expenditures support impoverished families to effectively raise their
children to stay away from crime, perhaps by allowing parents to spend
more time with their children.  Still, there remains significant
unexplained between-state heterogeneity in the effect of abortion on
murder, which could be due to effect modification driven by an
unobserved state-level moderator.  We do not discuss this point of
causal mechanisms any further except to note that this opens the door
to further research into the relationship between abortion rates and
the incidence of crime.

\subsection{Diagnostics for assumption of linear exposure effects}
\label{sec:diagnostics}

A key assumption of our model is that the exposure effect is linear in
$z$, conditional on the other covariates.  Here we introduce a model
check heuristic to diagnose the validity of this assumption.  To
begin, first note that we can rewrite the model (\ref{eq:model}) in
the form
\begin{align}
  \label{eq:linear-form}
  y - \mu(x) &= \tau(x) \cdot z + \varepsilon. 
\end{align}

Suppose that we partition observations into $J$ disjoint groups $g_j$
with similar estimated treatment effects, so that
$\tau(x_i) \approx \tau(x_{i'})$ for two observations $i, i' \in g_j$.
Let $\bar \tau_{g_j} = |g_j|^{-1}\sum_{i\in g_j}^{} \hat \tau_i $
denote the posterior mean for the group-level average treatment effect
within group $g_j$, where $\hat \tau_i \equiv \hat \tau(x_i)$ is the
posterior mean of the treatment effect for observation $i$. From
Eq.~\eqref{eq:linear-form}, we can see that the linearity assumption
implies
\begin{align}
  \label{eq:lin-group}
  \E [y_i - \hat \mu(x_i)] &\approx \bar \tau_{g_j} \cdot z_i,
                   \quad i \in g_j
\end{align}
where $\hat \mu$ is the posterior mean of the control function $\mu$.
That is, if the linearity assumption holds, then within a group $g_j$
of observations with similar estimated treatment effects $\hat \tau_i$
the relationship between the exposure $z_i$ and the partial residuals
$y_i - \hat \mu(x_i)$ is approximately linear with intercept 0 and the
slope given by $\bar \tau_{g_j}$.

With this in mind, we can deduce a simple and readily available way to
assess the linearity assumption from the data and the fitted model.
This may be done by plotting the partial residuals
$y_i - \hat \mu(x_i)$ versus the exposure $z_i$ for all units $i$ with
a group $g_j$.  The resulting scatterplot should show a linear
relationship for all groups.  If not, this suggests that the linearity
assumption is invalid, and the model estimates are suspect.  Note here
that the covariate vectors $x_i$ need not be similar within groups,
rather only their estimated treatment effects.

\begin{figure}[t!]
  \centering
  \includegraphics[width=0.75\textwidth]{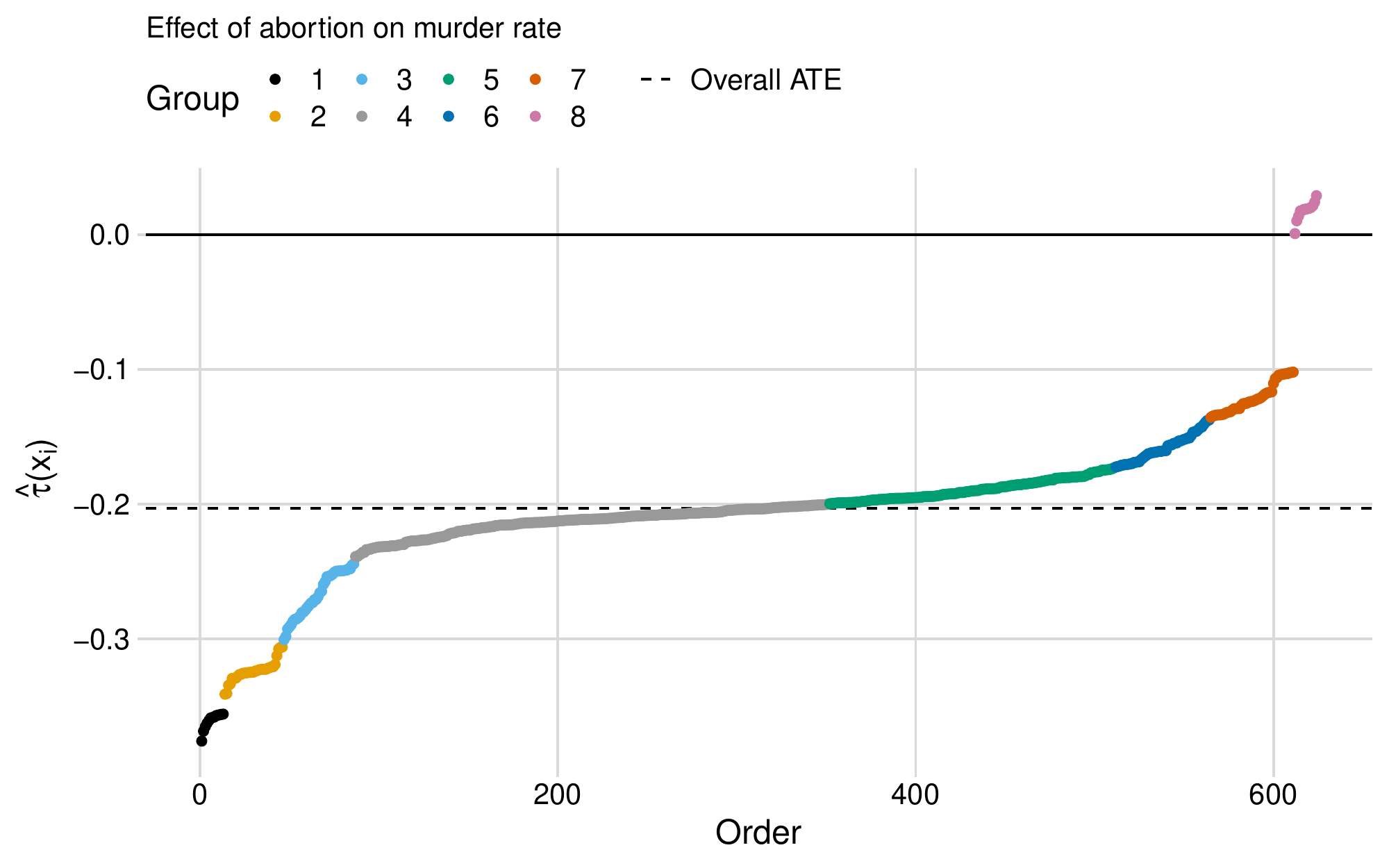} \\
  \includegraphics[width=0.75\textwidth]{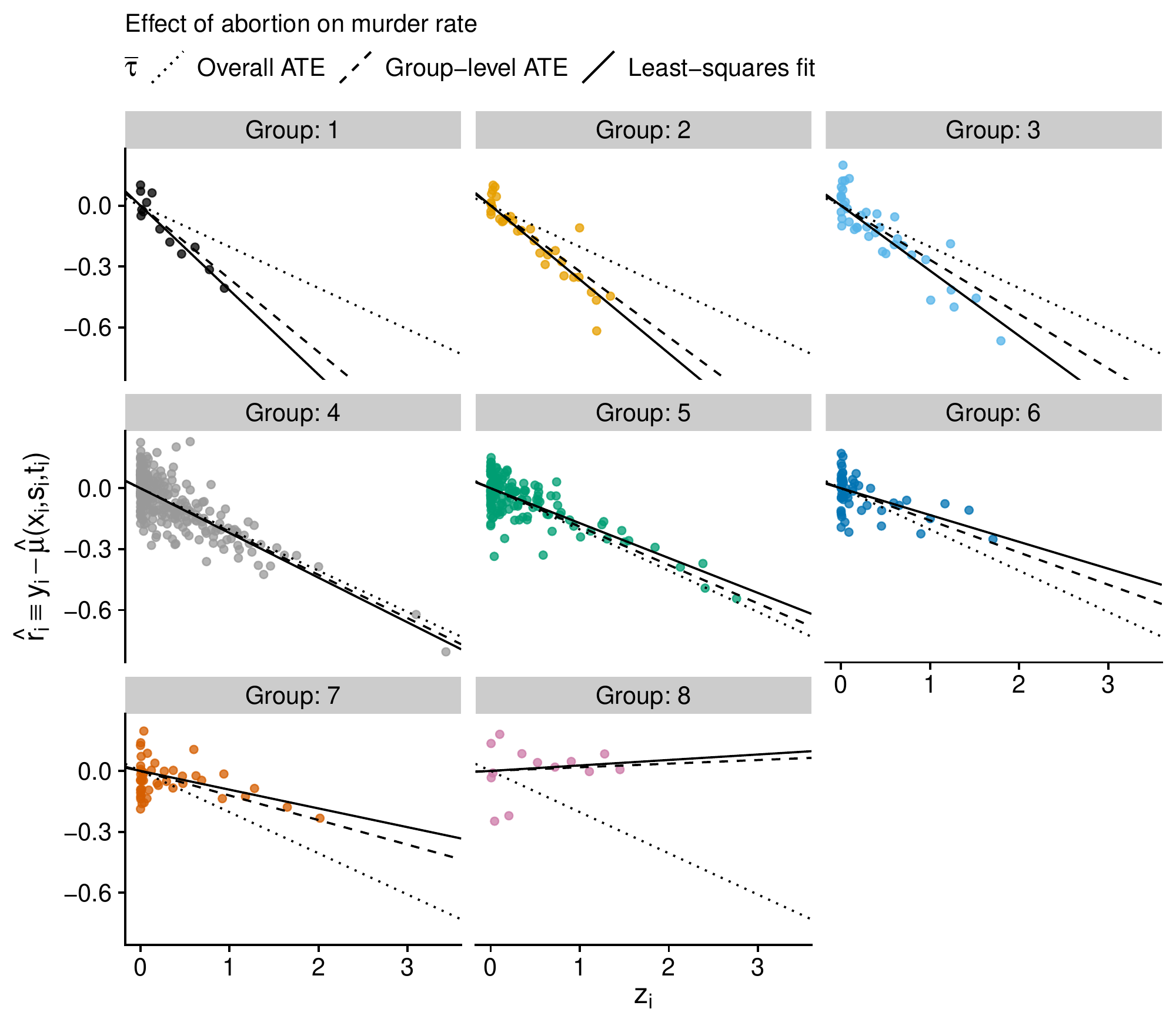}
  \caption{\label{fig:levitt-tau-clust} \textit{Top:} We create groups
    of observations with similar estimated exposure effects by
    performing hierarchical clustering on their posterior mean
    treatment effects $\hat \tau_i = \hat \tau(x_i, s_i, t_i)$.  We
    cut the dendrogram at a height equal to the standard deviation of
    these treatment effects, resulting in eight groups.
    \textit{Bottom:} We assess the linearity assumption by plotting
    the partial residuals $\hat r_i \equiv y_i - \hat \mu(x_i)$
    against the effective abortion rate $z_i$ for each observation,
    where $\hat \mu(x_i)$ is the posterior mean for the control
    function $\mu$ for each of the constructed groups.  For the
    scatterplot in each group, we also show three different fitting
    lines, (i) the overall ATE,
    $\bar \tau \equiv N^{-1}\sum_{i=1}^{N} \hat \tau_i$, (ii) the
    group-level ATE
    $\bar \tau_{g_j}= |g_j|^{-1}\sum_{i\in g_j}^{N} \hat \tau_i$, and
    (iii) the least-squares fit of the partial residuals $\hat r_i$ on
    the exposure $z_i$.}
\end{figure}

Now we apply this diagnostic heuristic
to our analysis of the Donahue and Levitt data. We calculate the
posterior mean treatment effects for all units,
$\hat \tau_i \equiv \hat \tau(x_i, s_i, t_i)$, and then form groups of
observations with similar treatment effects by performing hierarchical
clustering on the vector of $\hat \tau_i$ values.  We cut the
dendrogram at a height equal to the sample standard deviation of
$\hat \tau_i$, resulting in $J=8$ total groups $g_j$.  The top panel
of Figure~\ref{fig:levitt-tau-clust} orders the estimated treatment
effects $\hat \tau_i$ along the $x$-axis and shows the resulting eight
groups of observations.

The bottom panel of Figure~\ref{fig:levitt-tau-clust} shows
scatterplots of the partial residuals
$\hat r_i \equiv y_i - \hat \mu(x_i, s_i, t_i)$ versus the effective
abortion rate $z_i$ for all eight groups of observations.  For each
group $g_j$, we also show three different fitting lines, (i) the
overall ATE, $\bar \tau \equiv N^{-1}\sum_{i=1}^{N} \hat \tau_i$, (ii) the
group-level ATE $\bar \tau_{g_j}$, and (iii) the least-squares fit of
the partial residuals $\hat r_i$ on the treatment $z_i$.  Each group's
scatter plot shows a high degree of linearity.  This evidence is
supportive of our assumption that the causal effect of abortion on the
log-murder rate is linear once we condition on observed control
variables, state, and year.  Interestingly, we can also see a
strikingly clear partial pooling effect: each group level ATE
$\bar \tau_{g_j}$ is shrunk toward the overall ATE $\bar \tau$
compared to the group-specific least-squares line.

\begin{figure}[t!]
  \centering 
  \includegraphics[width=0.7\textwidth]{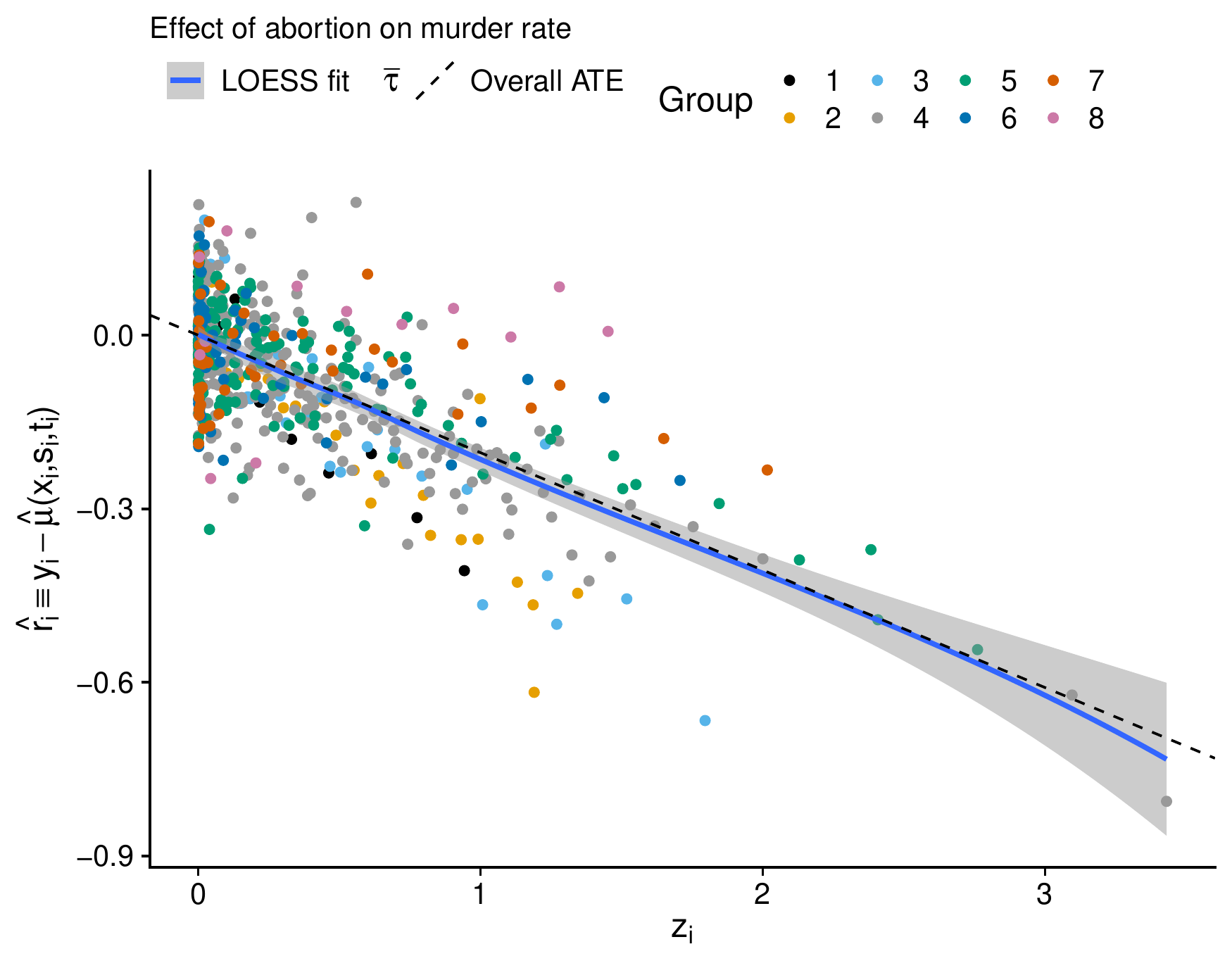}
  \caption{\label{fig:cluster-scatter-all} Plot of partial residuals
    $y_i - \mu(x_i)$ vs. treatment $z_i$ for every observation.  A
    loess smoother fit to these data shows a fit remarkably close to
    linear, and well approximated by the linear fit implied by the
    overall average treatment effect.  }
\end{figure}

There remains a possibility assumption of locally linear exposure
effects (i.e., linear within groups) could disguise a partial
exposure-response curve which is globally nonlinear.  Therefore, in
Figure~\ref{fig:cluster-scatter-all} we show a scatterplot of the
partial residuals $\hat r_i$ against the exposure $z_i$ concatenated
for all groups of observations.  A loess smoother fit to these data
shows a fit which is remarkably close to linear, and well approximated
by the linear fit implied by the overall average treatment effect
$\bar \tau$.

We feel confident in the linearity assumption as applied to analyzing
the effect of abortion on the murder rate.  We show in the supplement
that these model checks perform similarly well when using our model on
the violent crime and property crime data.  In the next section, we
provide a simulated example where our model check is able to detect a
scenario where this linearity assumption does not hold by
construction.


\section{Experimental results}
\label{sec:experimental-results}

In this section we investigate how our model performs when the
assumption of linear exposure effects is violated, and how our model
diagnostic procedure can detect this violation.  We generate outcome
data via the following mechanism:
\begin{align}\label{eq:dgp}
  \begin{split}
      y &= \mu(x) + h(z) + \epsilon, \quad \epsilon \sim \N(0, 0.5^2)  \\
  \mu(x) &= x/2 \\ 
  h(z) &= (z + 1)^2/2 + 1/4
  \end{split}
\end{align}
with a univariate $x$ and exposure $z$.  With this data generating
mechanism, two of the assumptions of our model in \eqref{eq:model} are
violated.  First, whereas model \eqref{eq:model} assumes that,
conditional on the covariates, the effect of $z$ on $y$ is linear,
here the exposure effect is in fact quadratic. Second, model
\eqref{eq:model} assumes that the exposure effect is moderated by a
subset of the covariates, while here the exposure effect is not
affected by $x$.

Each $x$ is generated from a standard Gaussian, $x \sim \N(0, 1)$.  To
allow the inducement of confounding, the exposure variable is
generated conditional on $x$, first by generating
$(z' \mid x) \sim \N(b x - 1, 1)$ and then calculating $z$ by scaling
the vector of $z'$ values to have unit variance (so that $z$ has
constant marginal variance for all $b$).  The value of $b$ determines
the level of confounding of $x$ on $z$.  To explore the
characteristics of our model diagnostic procedure in depth, we
consider two cases of this data generating mechanism:
\begin{itemize}
\item \textit{Case 1}: Set $b=0$, so there is \textit{no
    confounding} of $z$ by $x$,
  and
\item \textit{Case 2}: Set $b=1$, so there is \textit{moderate
    confounding} of $z$ by $x$.
\end{itemize}
For both of these cases we generate $N=1000$ observations of $x$, $z$,
and $y$ as described above, and estimate the posterior for the model
in Eq.~\eqref{eq:model}.  We first present the resulting model
estimates, and then apply the model diagnostic procedure given in
Section~\ref{sec:diagnostics} to check the linearity assumption.

\begin{figure}[ht]
  \centering
  \includegraphics[width=0.9\textwidth]
  {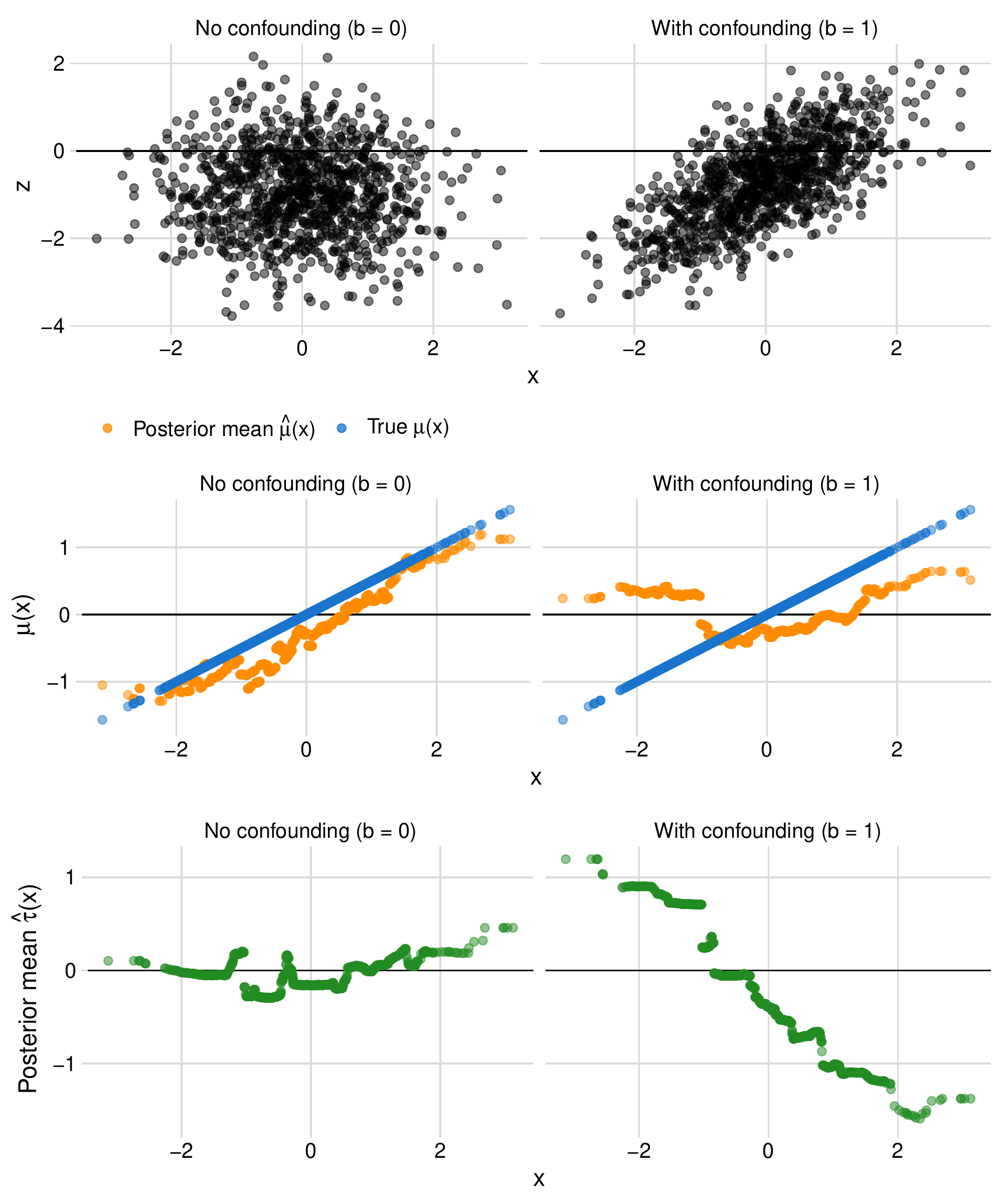}
    \caption{Simulated data and model estimates for the experiment in
      Section~\ref{sec:experimental-results}.  We consider two
      scenarios of confounding of $x$ on $z$: one where there is no
      confounding (left column), and one where there is a moderate
      level of confounding (right column).  \textit{Top}: a scatter
      plot of the control/moderating covariate $x$ and the exposure
      $z$. \textit{Middle}: The true control function $\mu(x)$ and the
      posterior mean control function $\hat \mu(x)$. \textit{Bottom}:
      The posterior mean for the effect moderating function $\tau$.
      Note that in the true data generating mechanism, the exposure
      effect is in fact not moderated by $x$.  As opposed to the case
      of no confoundingness, the confounded example misidentifies the
      control function and falsely detects a significant amount of
      effect heterogeneity across $x$. }
  \label{fig:sim-results1}
\end{figure}

Figure~\ref{fig:sim-results1} compares the results of the two
simulated examples, showing a scatterplot of the covariate and
exposure variables and the posterior means for the control and
moderating functions evaluated at each $x$.  In case 1 (no
confounding), the model detects little variation in the exposure
effect over $x$.  This is seen in the mostly flat posterior mean for
the moderating function $\hat\tau(x)$.  This function is centered on
0, which is the gradient of the true $h(z)$ function in
Eq.~\eqref{eq:dgp} averaged across the observed exposure vector $z$.
There is in fact variation in the exposure effect in the data
generating mechanism, but this variation depends on $z$ rather than
$x$.  In this sense, the model correctly identifies that there is no
moderation by $x$, though by construction it cannot detect that the
exposure effect is nonconstant in $z$.  Therefore, $\hat \tau(x)$ is
centered on 0 because the model averages over the gradient of $h(z)$
with respect to $z$.  Finally, because the model does not falsely
attribute variation in the effect of $z$ to $x$, it is able to closely
capture the control function $\mu(x)$.

However, in case 2 (moderate confounding) the model mistakenly detects
moderation of the exposure effect by $x$, with the downward-sloping
$\hat \tau(x)$ function indicating that increased $x$ lowers the
exposure effect, with the exposure effect turning from positive to
negative near $x=-1$.  We can intuit how this finding is a consequence
of the presence of confounding and the incorrect assumption of linear
exposure effects.  Here $x$ and $z$ are collinear, so lower values of
$x$ are associated with lower values of $z$, where the gradient of
$h(z)$ in the true data generating mechanism \eqref{eq:dgp} is
positive for $z < -1$.  Likewise, higher values of $x$ are associated
with higher values of $z$, where the gradient of $h(z)$ is negative
for $z > -1$.  In other words, $x$ acts as a proxy for $z$, and so $x$
is predictive of where in the quadratic $h(z)$ function a particular
observation lies.  Because the model we use assumes that the exposure
effect is both linear and moderated by $x$, this phenomenon manifests
itself in the downward-sloping estimated moderating function
$\hat \tau(x)$.  As a further result of this misidentification, the
estimate for the control function deviates rather significantly from
the truth.

\begin{figure}[ht]
  \centering
  \includegraphics[width=0.99\textwidth]{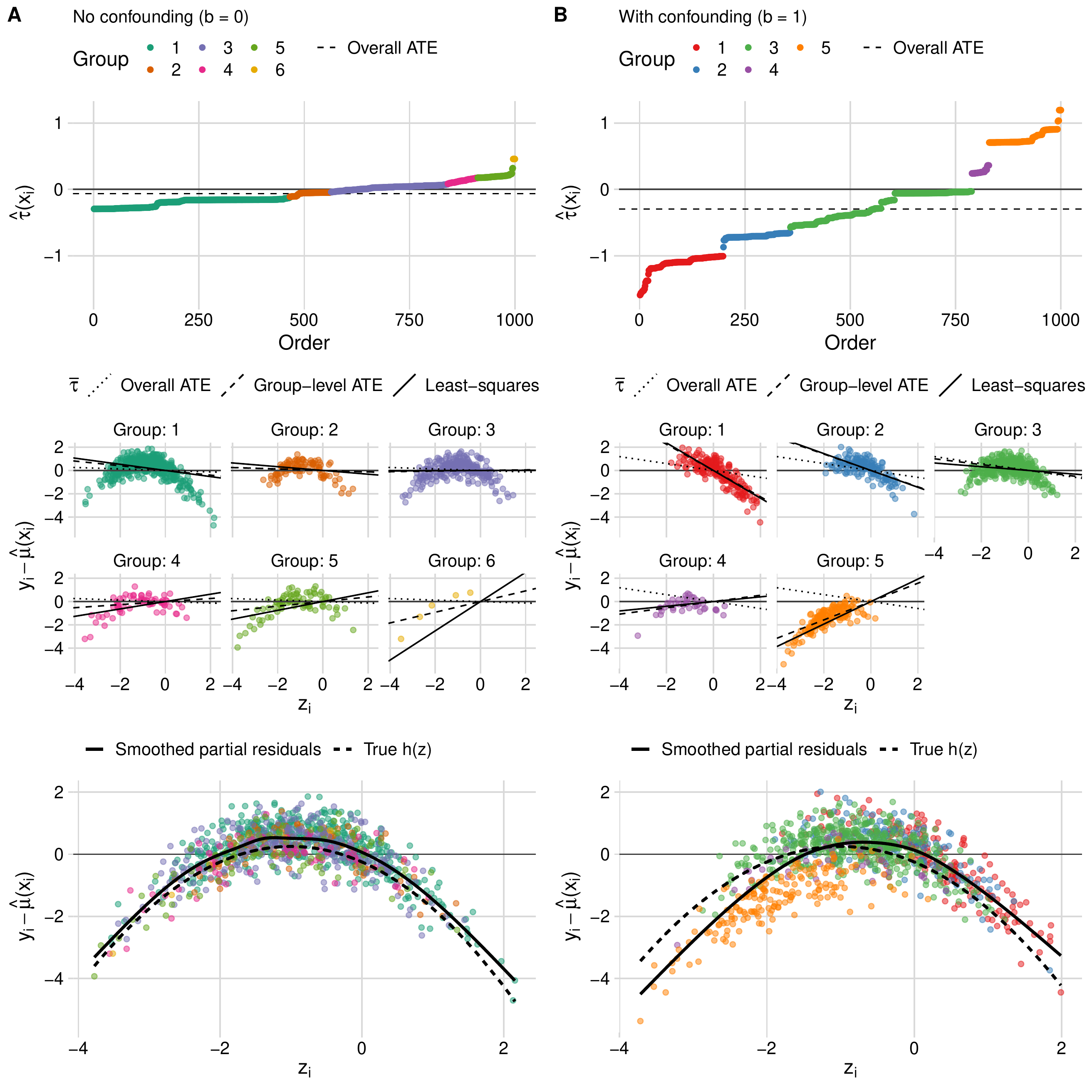}
  \caption{Model diagnostic procedure applied to the simulation
    experiment in Section~\ref{sec:experimental-results}, for both the
    unconfounded and confounded cases (left and right columns,
    respectively).  \textit{Top}: the posterior mean individual
    treatment effects $\hat \tau_i$ are arranged and grouped using
    hierarchical clustering. \textit{Middle}: A plot of the partial
    residuals $y_i - \hat \mu(x_i)$ against the exposure $z_i$,
    faceted by the grouping of observations in the top panel.  If the
    linearity assumption holds, then the relationship within each
    group should be approximately linear.  However, for both examples,
    we see clear violations of the linearity assumption, quite
    noticeable in the unconfounded case, and also visible in some
    groups for the confounded case (notably groups 3 and
    5). \textit{Bottom}: A plot of the partial residuals against the
    exposure combined for all observations.  Even though the
    assumption of local linearity is violated, our model appears to
    closely capture this particular nonlinear partial exposure effect
    function. }
  \label{fig:sim-results-scatter}
\end{figure}

It is important to note that, for both cases, the estimated (linear)
exposure effects are ultimately unreliable because the linearity
assumption is incorrect.  In fact, our model diagnostic procedure is
able to detect this fact.  Figure~\ref{fig:sim-results-scatter}
demonstrates the model diagnostic procedure, outlined for the main
application in Section \ref{sec:diagnostics}, as applied to the two
simulation examples.  As we did for the application, for both cases 1
and 2 we construct groups of observations $g_j$ with similar estimated
treatment effects by performing hierarchical clustering on the values
of the posterior mean for the treatment effects $\hat \tau(x_i)$
(shown in the top panel of Figure~\ref{fig:sim-results-scatter}).

Then we plot the partial residuals
$\hat r_i \equiv y_i - \hat \mu(x_i)$ against the exposure $z_i$ to
check for a linear relationship within each group of observations
constructed in the first step (middle panel of
Figure~\ref{fig:sim-results-scatter}).  That is, if the linearity
assumption holds, then within each group $g_j$ the relationship
between $\hat r_i$ and $z_i$ should be approximately linear with
intercept 0 and slope
$\bar \tau_{g_j} \equiv |g_j|^{-1}\sum_{i \in g_j}^{} \hat \tau(x_i)$.
For both simulations, we can notice telltale signs of departures from
the our model's core linearity assumption.  For case 1 (no
confounding), the scatterplot for each group has a high degree of
nonlinear curvature.  This is due to the fact that there was little
detected effect heterogeneity in $x$; the treatment effect
$\hat \tau(x)$ is close to 0 for each $x$, meaning that each group of
observations contains values of the exposure $z$ distributed fairly
evenly among the parabolic shape of the $h(z)$ function.

Meanwhile, for case 2 (moderate confounding), the violation of the
linearity assumption is also readily apparent, though slightly less
obvious compared to case 1.  As discussed previously, the presence of
confounding produces erroneously large estimates of effect moderation
by $x$.  The scatterplots for each group show how the parabolic shape
of the $h(z)$ function is cut into regions where its gradient is
approximately constant.  This reinforces our previous remark that $x$
serves as a proxy for where an observation is situated in the $h(z)$
function because of the collinearity between $x$ and $z$.  Even so,
there is some remaining nonlinearity present in these scatterplots,
most clearly seen for groups 3 and 5 for case 2.

Taken together, it is clear that our model diagnostic procedure can
reliably detect the existence of nonlinearity in the true exposure
effect, even with the presence of confounding.  For an applied data
analysis, similar diagnostic results of this kind would convey that
the model's key linearity assumption is violated, and so an alternate
model specification may be necessary.  However, for our main
application in Section~\ref{sec:analys-donah-levitt}, this model
diagnostic heuristic was unable to detect any significant deviations
from linear exposure effects.  In our view, this lends credence to the
legitimacy of maintaining the linearity assumption for our analysis of
the effect of abortion on crime rates.

As a final mode of investigation, we plot the partial residuals
against the exposure once more, this time for all observations
combined (bottom panel of Figure~\ref{fig:sim-results-scatter}).  A
loess smoother is fit to these data, and for comparison, we also show
the true exposure effect function $h(z)$ used in the data generating
process.  Interestingly, the loess curve for the scatterplot of both
simulation cases shows a close fit to $h(z)$.  The loess fit shows
slightly more deviation from $h(z)$ because the estimated control
function did not closely match the truth.  This evidence insinuates
our model is able to estimate the partial exposure effect, even if the
core model assumption of linear exposure effects is violated.  This
estimate of the partial exposure effect is more robust when there is
no confounding, and is likely helped by the fact that there is
stronger signal in the exposure than in the control function.


\section{Discussion}
\label{sec:discussion}

In this paper, we have introduced a model which may be used to
estimate heterogeneous effects of continuous treatments and exposures.
We decouple the regularization of the control function, which
encapsulates the role of control variables, and the moderating
function, which describes effect modification by moderating
covariates.  Crucially, our model specification is a priori agnostic
to the exact role of control covariates in the outcome model, allowing
for nonlinearities and interactive trends to exist.  This reduces
researcher degrees of freedom in applied data analyses, thus improving
the robustness and reproducibility in reported results.

We find that there is strong support for the existence of a negative
causal effect of abortion on murder, violent crime, and property
crime, consistent with the findings of \cite{Levitt}.  Our model
diagnostic check lends credence to our main parametric assumption of
linear exposure effects conditional on the value of moderators.
Through the use of posterior summarization, we discover suggestive
evidence that this causal relationship is tempered when there is
larger state expenditures to provide monetary assistance to poor
families with dependent children.  Still, there is a significant
degree of remaining unexplained heterogeneity in the magnitude of this
effect between the states, and explaining this variation provides an
opening to further study of the dynamic between abortion and crime.

We emphasize that this exploration of the effect heterogeneity
resulting from our model estimates did not require us to fit multiple
models, e.g. refitting a model which interacts the exposure of
abortion with state dummy variables.  Repeated uses of the outcome
data in such a way induces problems of multiplicity which are likely
intractable.  This is one more way in which our approach preserves
reproducibility.












\renewcommand{\thesubsection}{\Alph{subsection}}
\appendix

\section{Analysis of effect of abortion on other crime outcomes}

\begin{figure}[ht]
  \centering
  \includegraphics[width=1\textwidth]{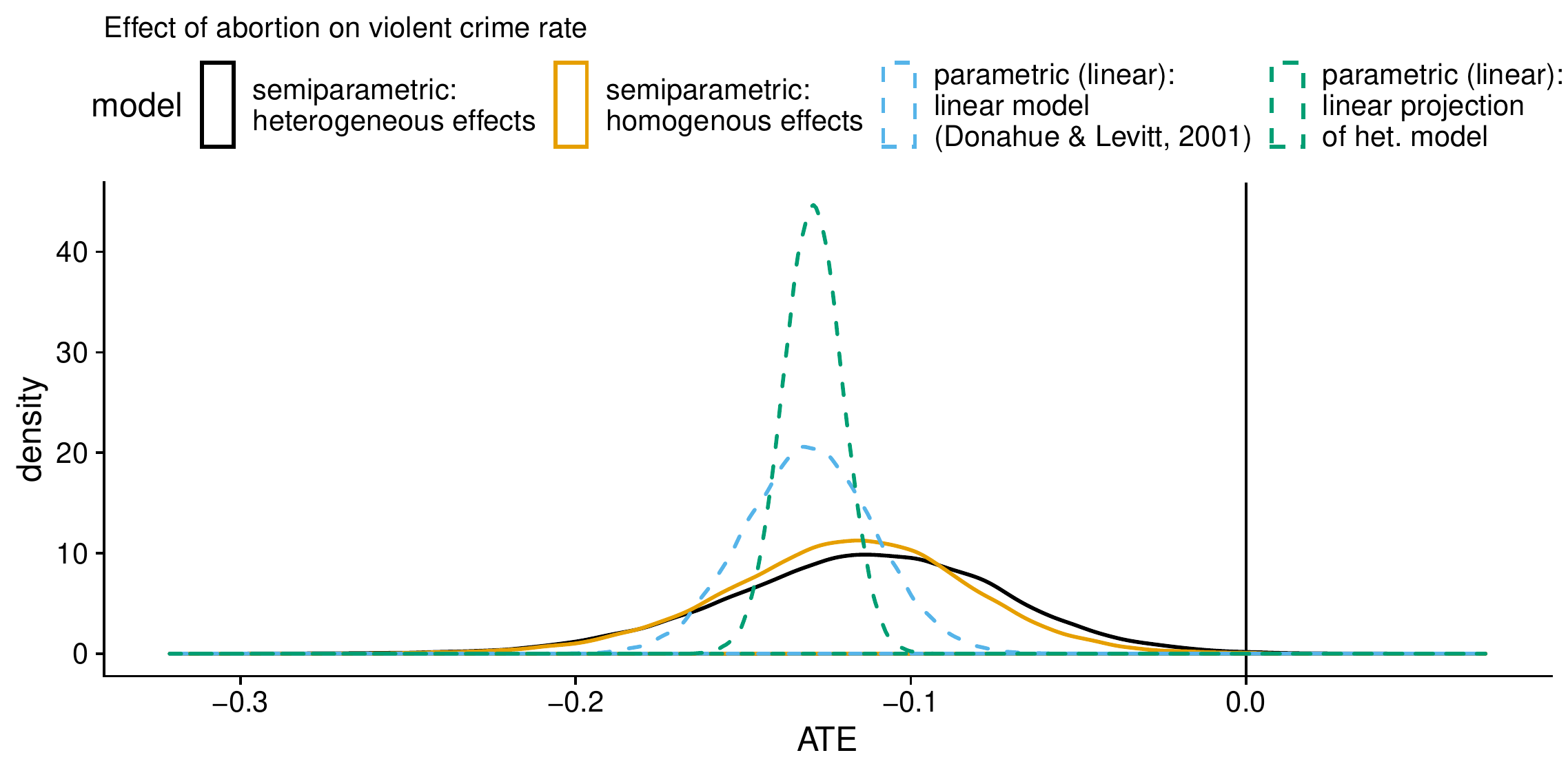}
  \caption{\label{fig:levitt-ate-v} Estimates of the average treatment
    effect (ATE) for the analysis of the Donahue and Levitt violent
    crime data.  }
\end{figure}

\begin{figure}[ht]
  \centering
  \includegraphics[width=1\textwidth]{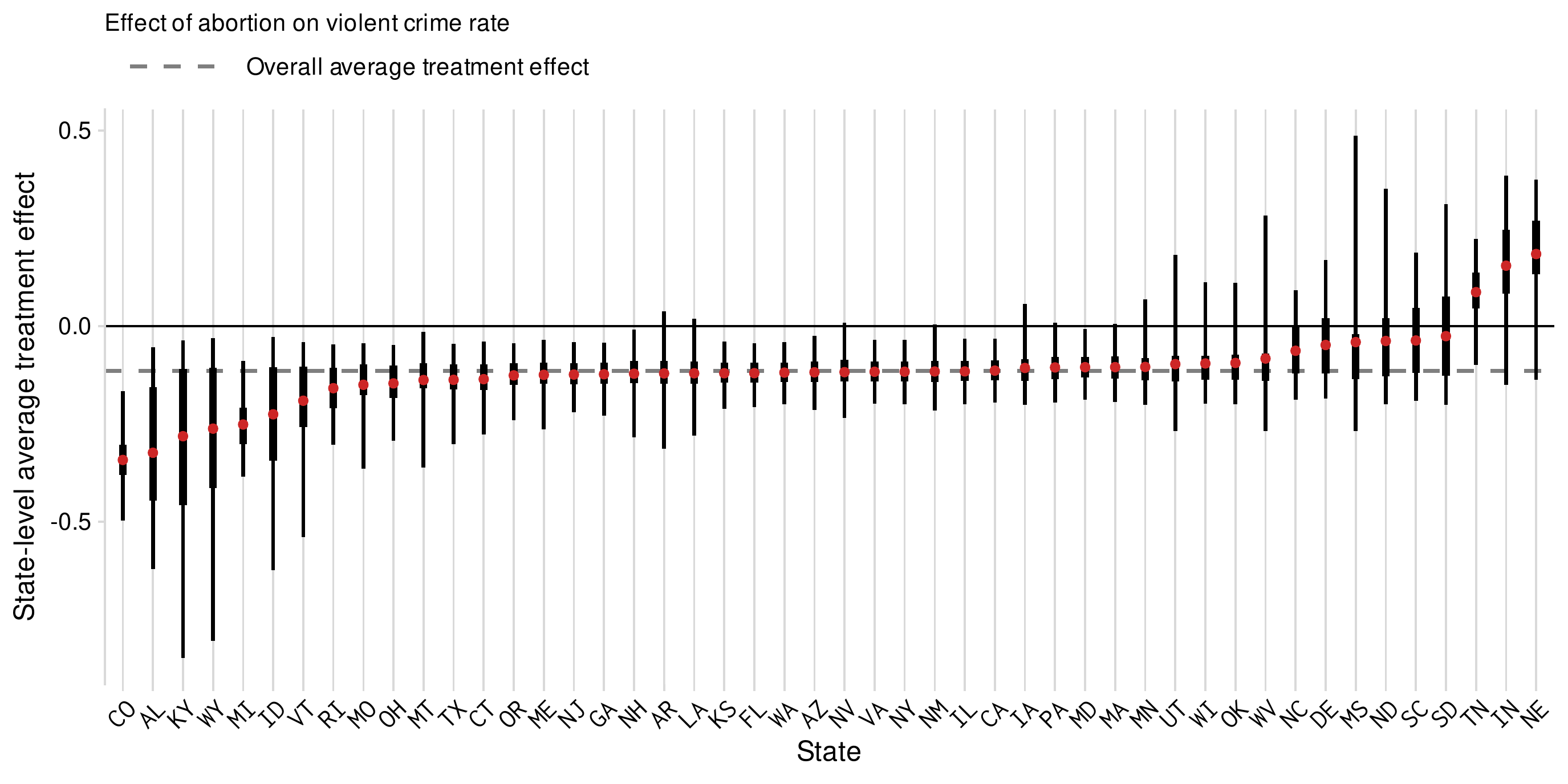}
  \caption{\label{fig:levitt-state-v} ATE estimate for each state for
    violent crime, along with equal-tailed 50\% and 95\% credible
    intervals. }
\end{figure}

\begin{figure}[t!]
  \centering
  \includegraphics[height=0.4\textheight]{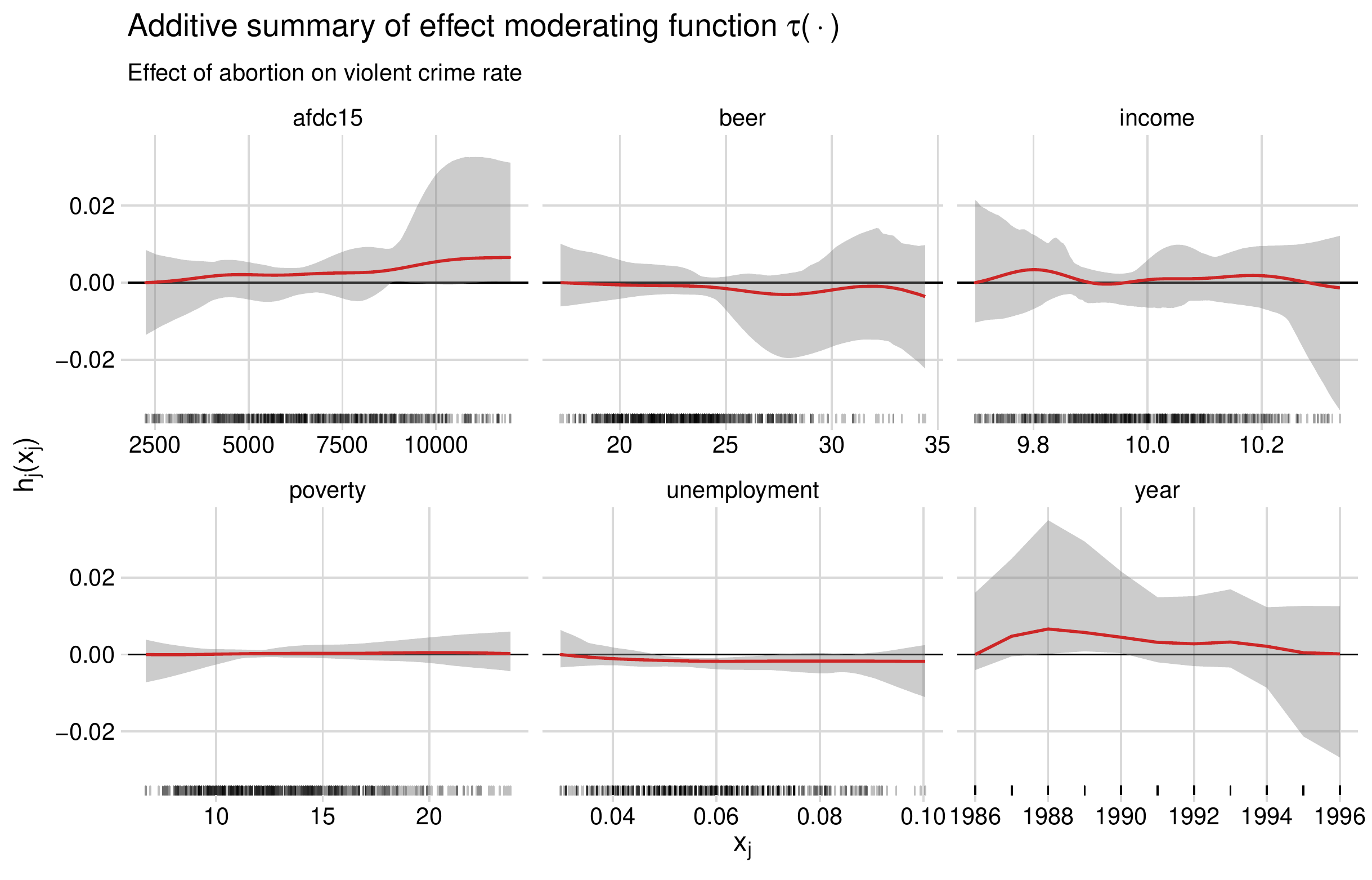} \\
  \caption{\label{fig:proj-tau-mu-v} Posterior summaries of the
    treatment modifying function $\tau(\cdot)$ for violent crime.  }
\end{figure}

\begin{figure}[ht!]
  \centering
  \includegraphics[width=0.75\textwidth]{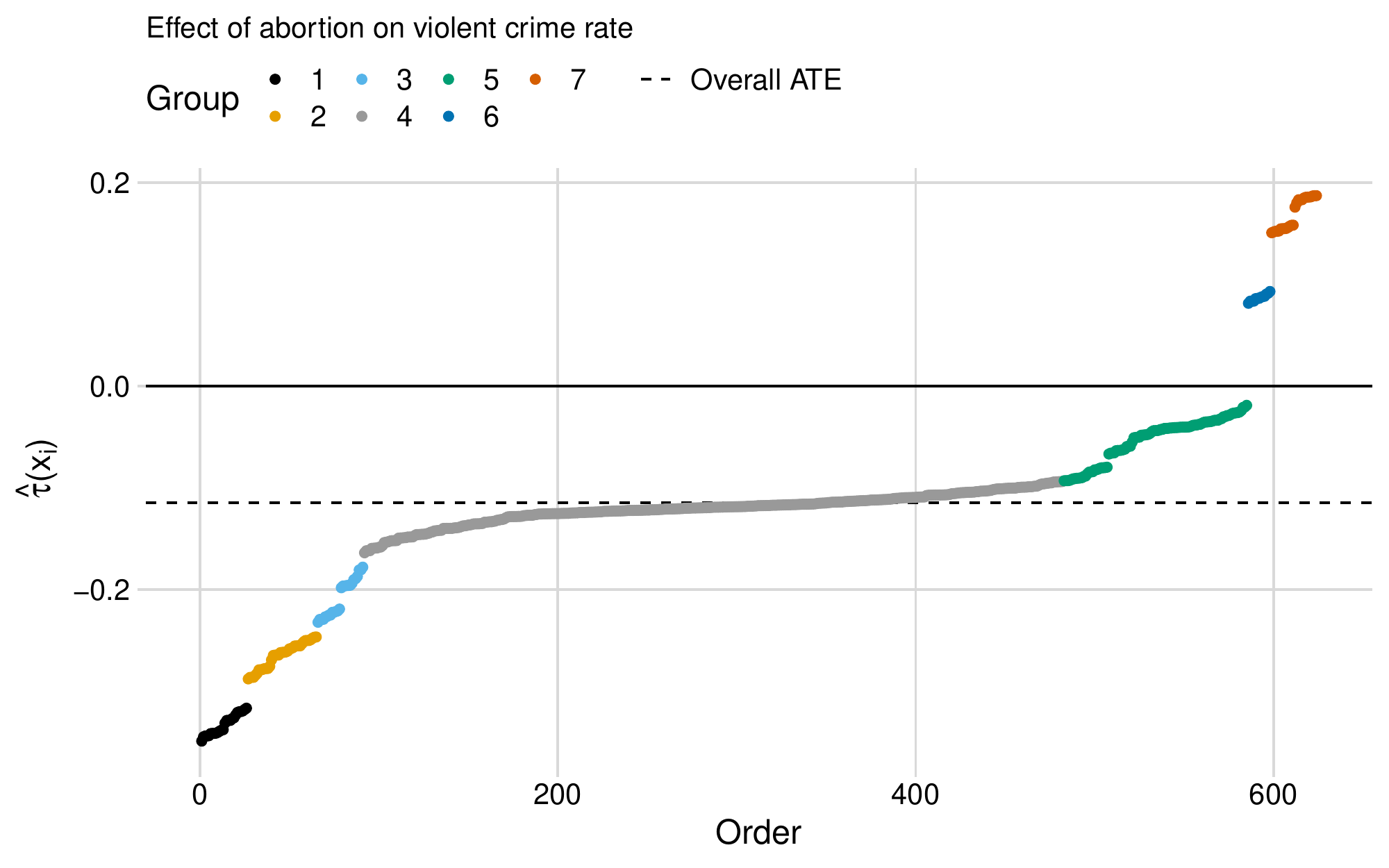} \\
  \includegraphics[width=0.75\textwidth]{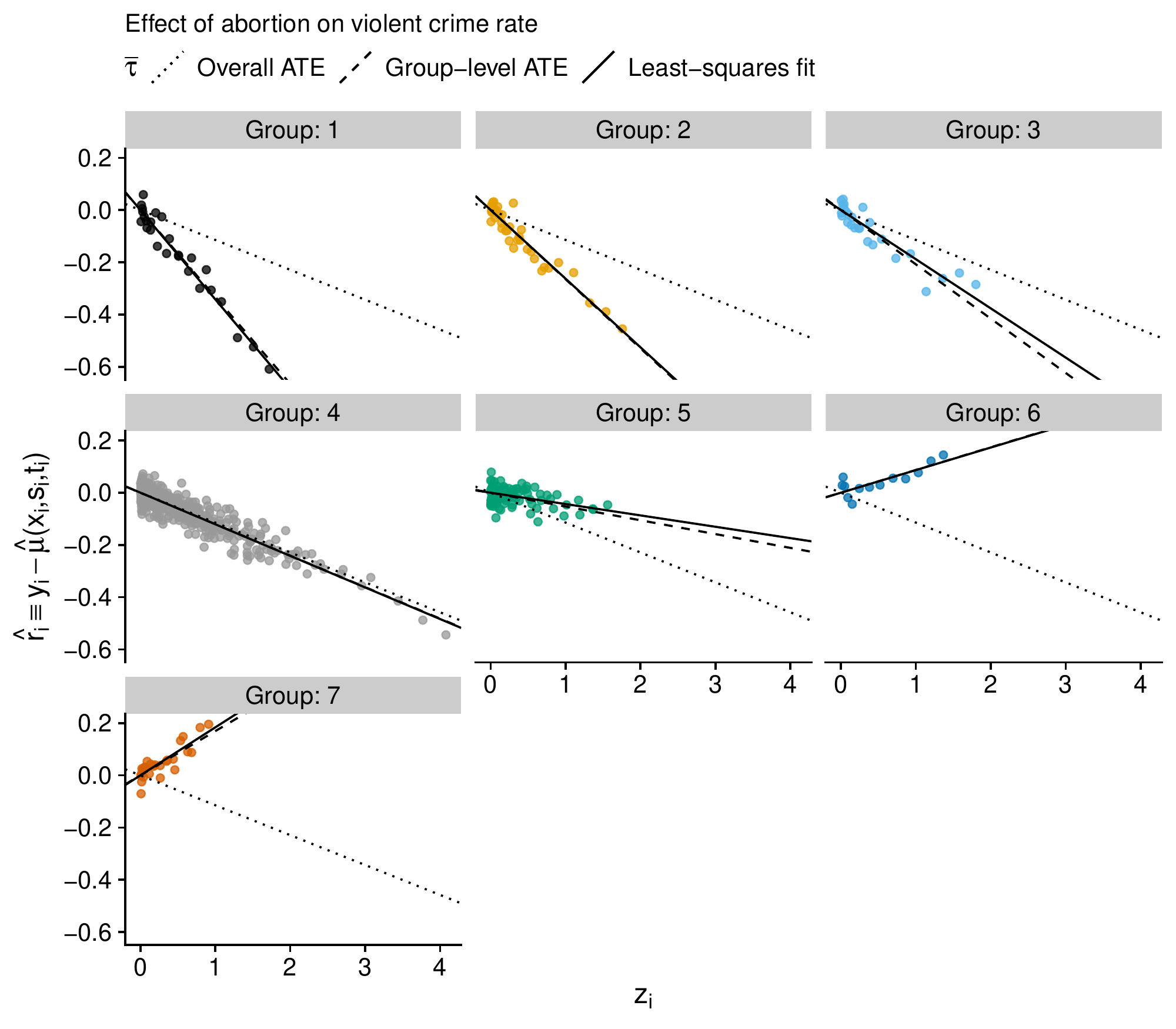}
  \caption{\label{fig:levitt-tau-clust-v} Diagnostics for the
    linearity assumption for the violent crime data. }
\end{figure}

\begin{figure}[ht!]
  \centering
  \includegraphics[width=0.7\textwidth]{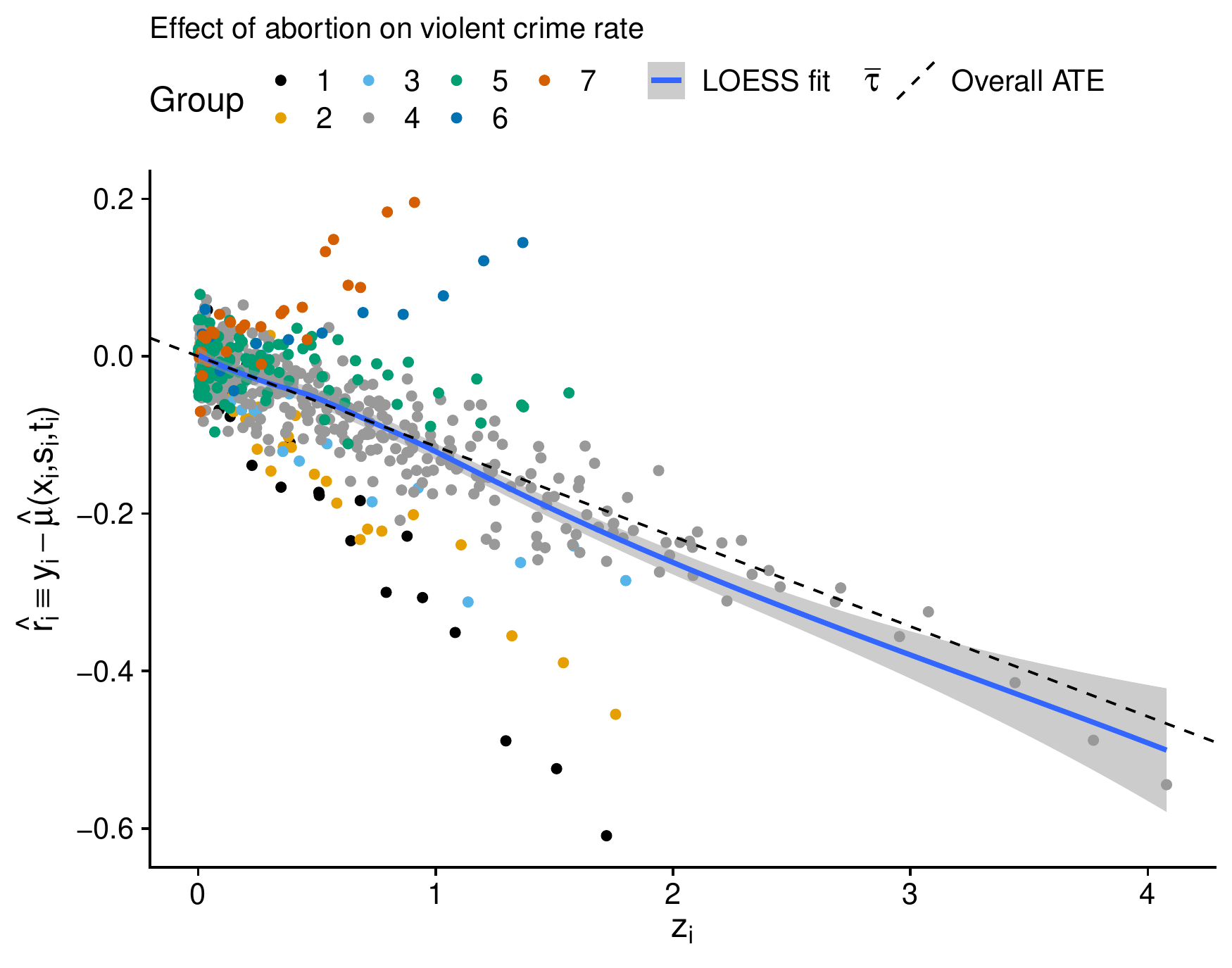}
  \caption{\label{fig:cluster-scatter-all-v} Plot of partial residuals
    $y_i - \mu(x_i)$ vs. treatment $z_i$ for every observation for the violent crime data. }
\end{figure}

\begin{figure}[ht]
  \centering
  \includegraphics[width=1\textwidth]{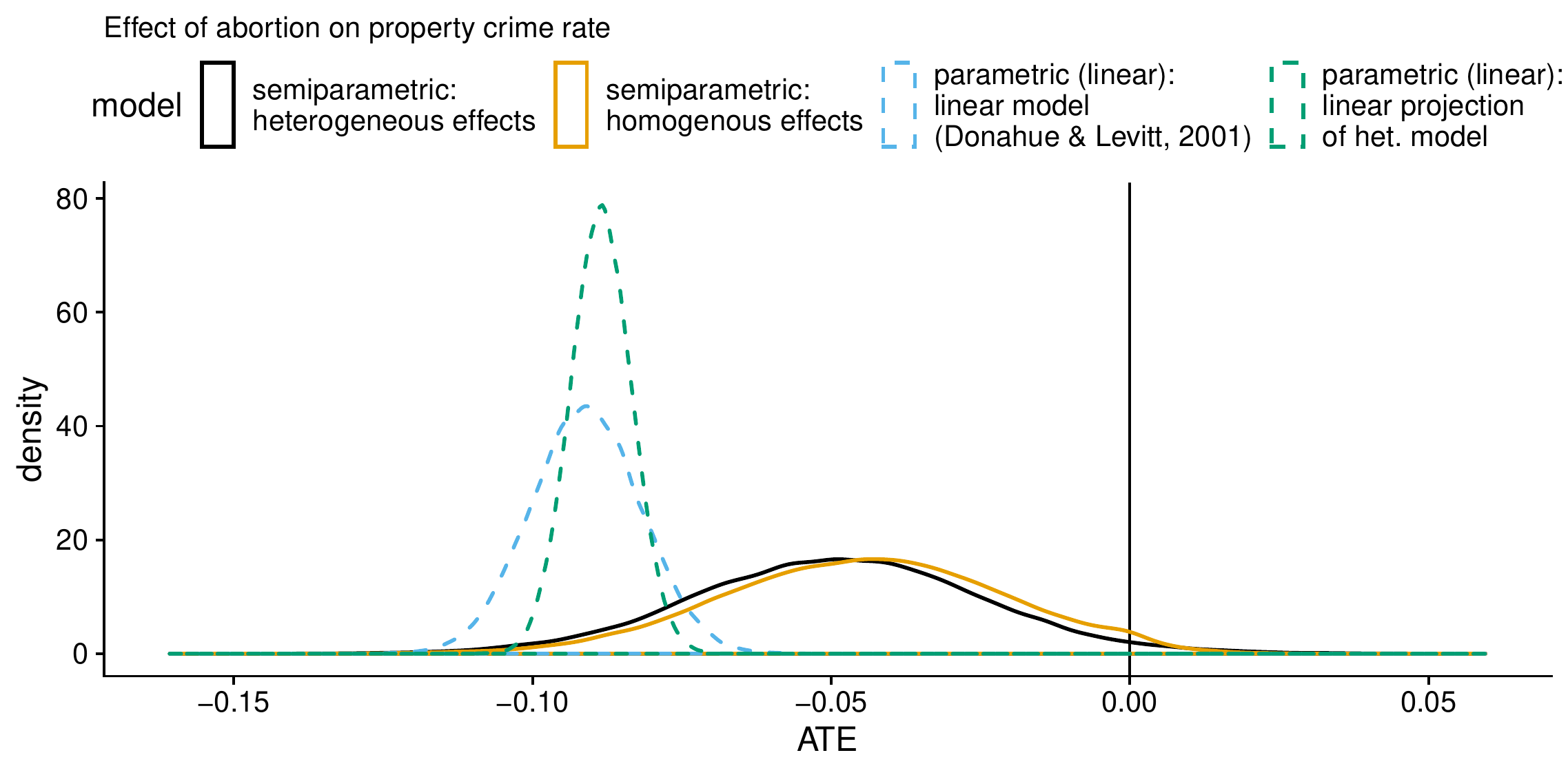}
  \caption{\label{fig:levitt-ate-p} Estimates of the average treatment
    effect (ATE) for the analysis of the Donahue and Levitt property
    crime data.  }
\end{figure}

\begin{figure}[ht]
  \centering
  \includegraphics[width=1\textwidth]{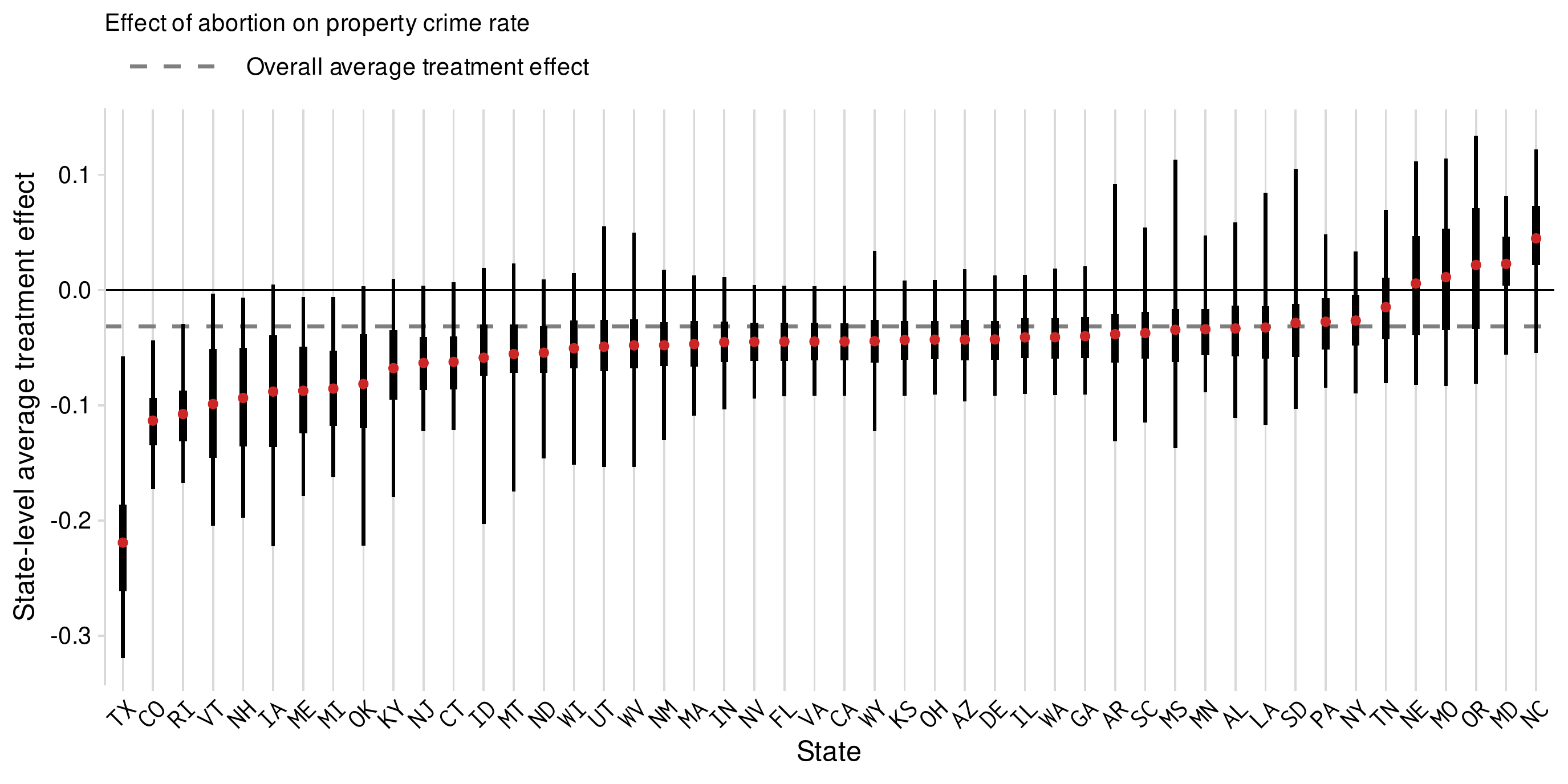}
  \caption{\label{fig:levitt-state-p} ATE estimate for each state for
    property crime, along with equal-tailed 50\% and 95\% credible
    intervals. }
\end{figure}

\begin{figure}[t!]
  \centering
  \includegraphics[height=0.4\textheight]{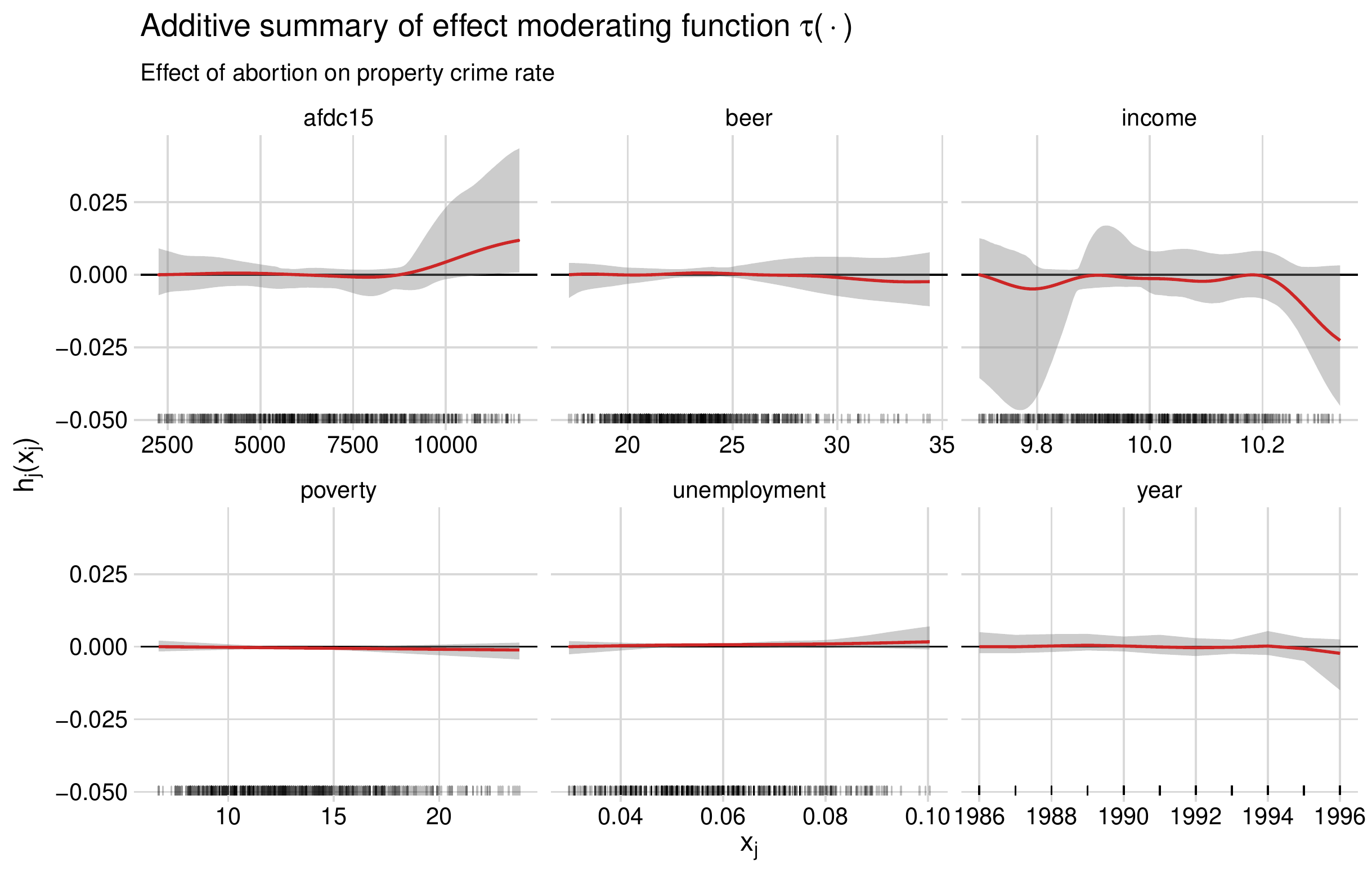} \\
  \includegraphics[width=0.35\textwidth]{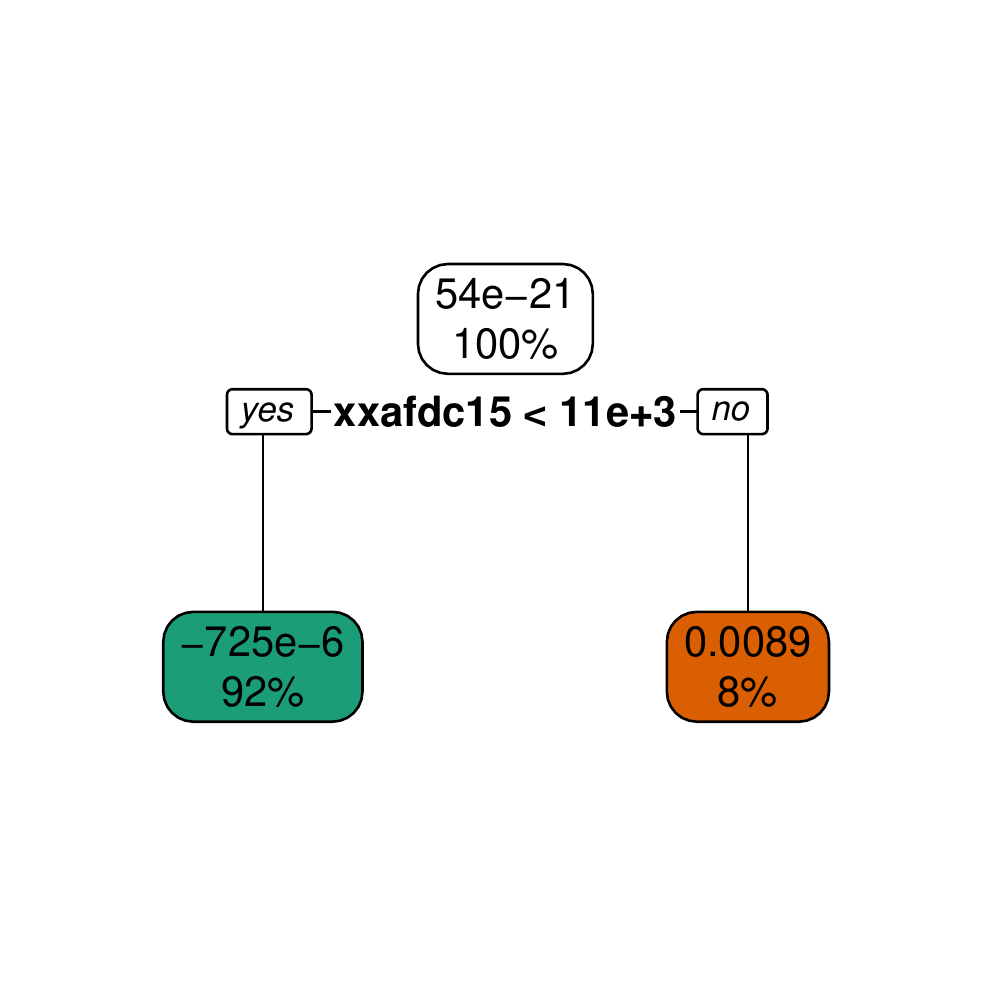}
  \includegraphics[width=0.64\textwidth]{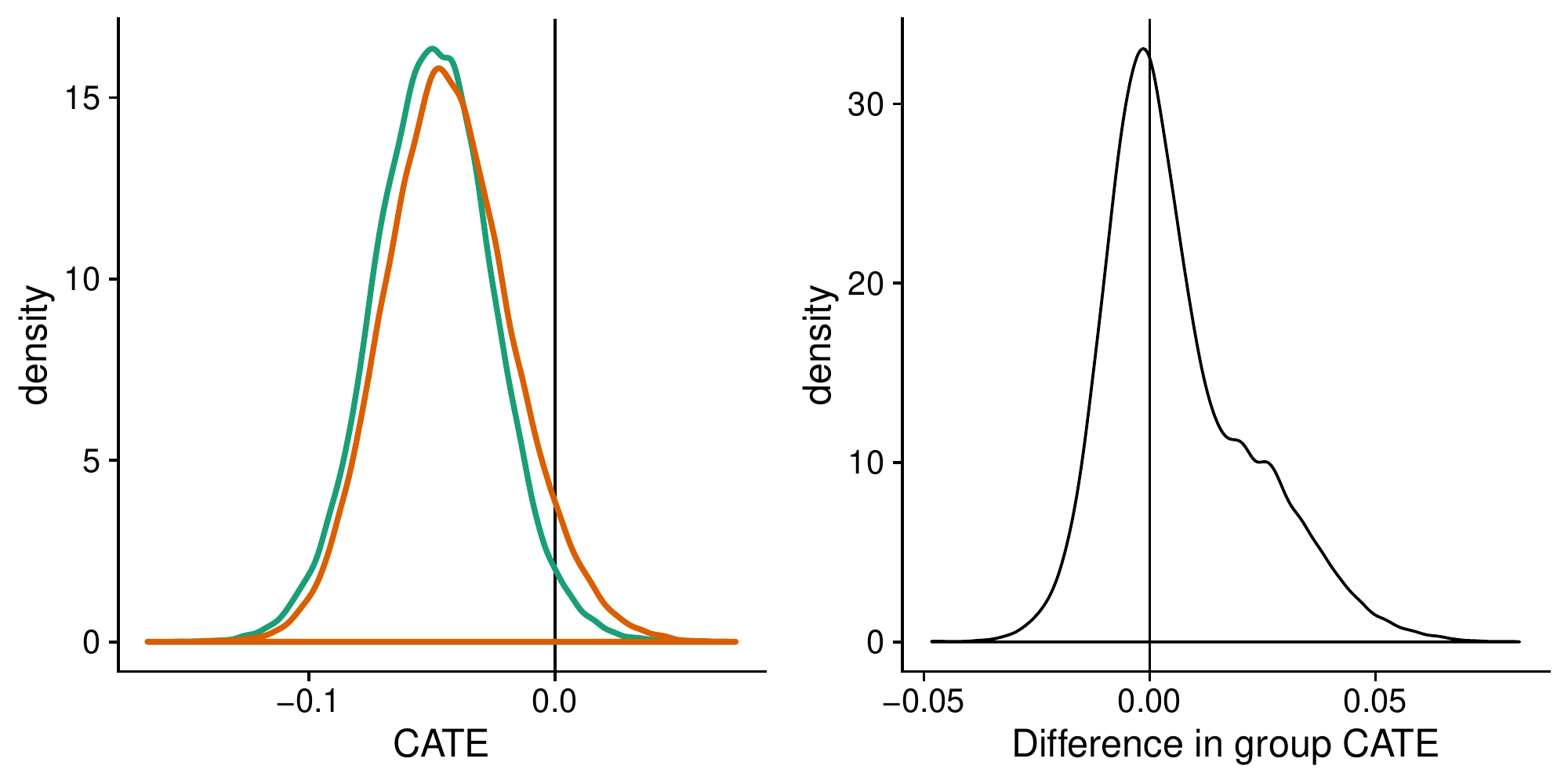}
  \caption{\label{fig:proj-tau-mu-p} Posterior summaries of the
    treatment modifying function $\tau(\cdot)$ for property crime.  }
\end{figure}

\begin{figure}[ht!]
  \centering
  \includegraphics[width=0.75\textwidth]{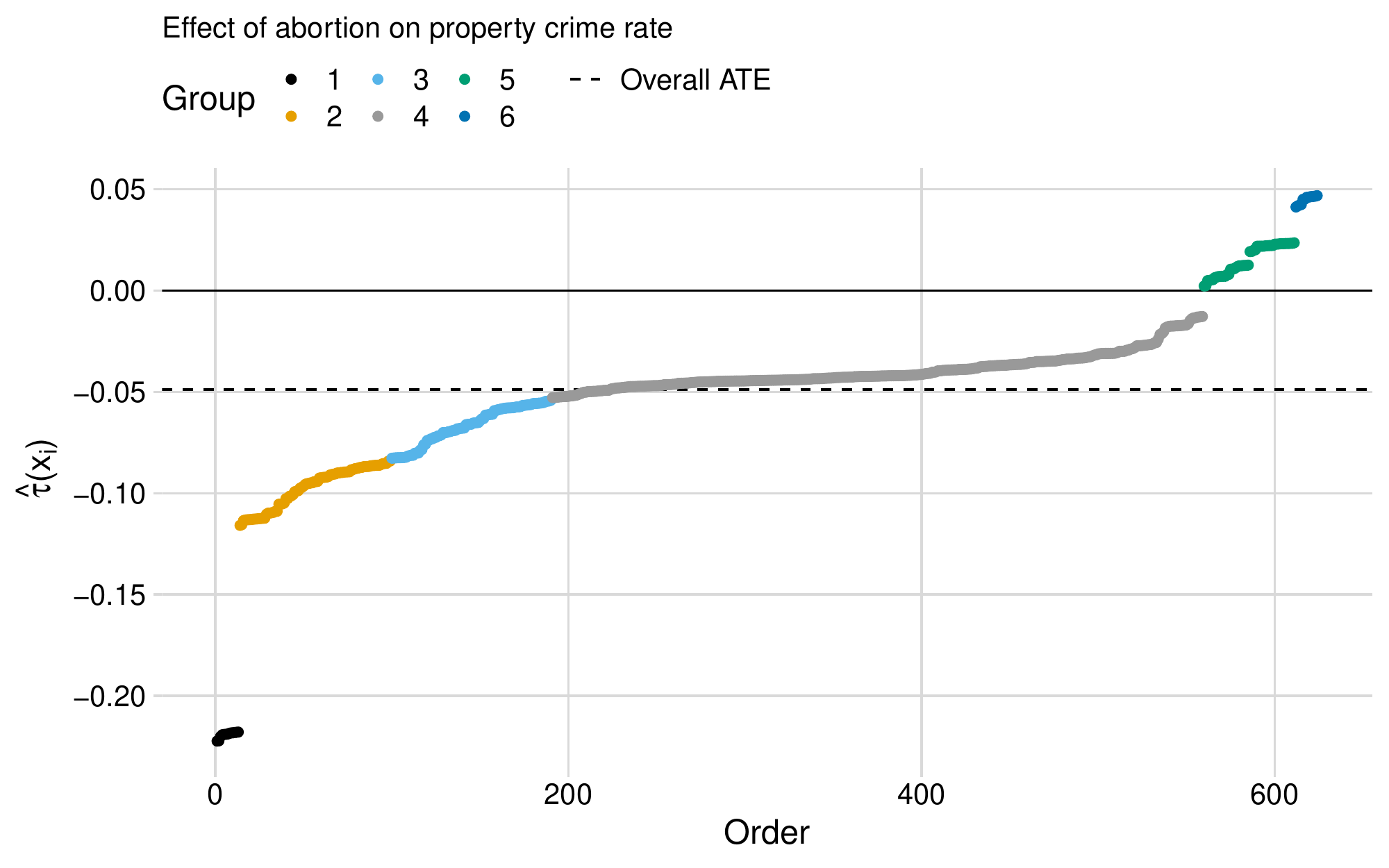} \\
  \includegraphics[width=0.75\textwidth]{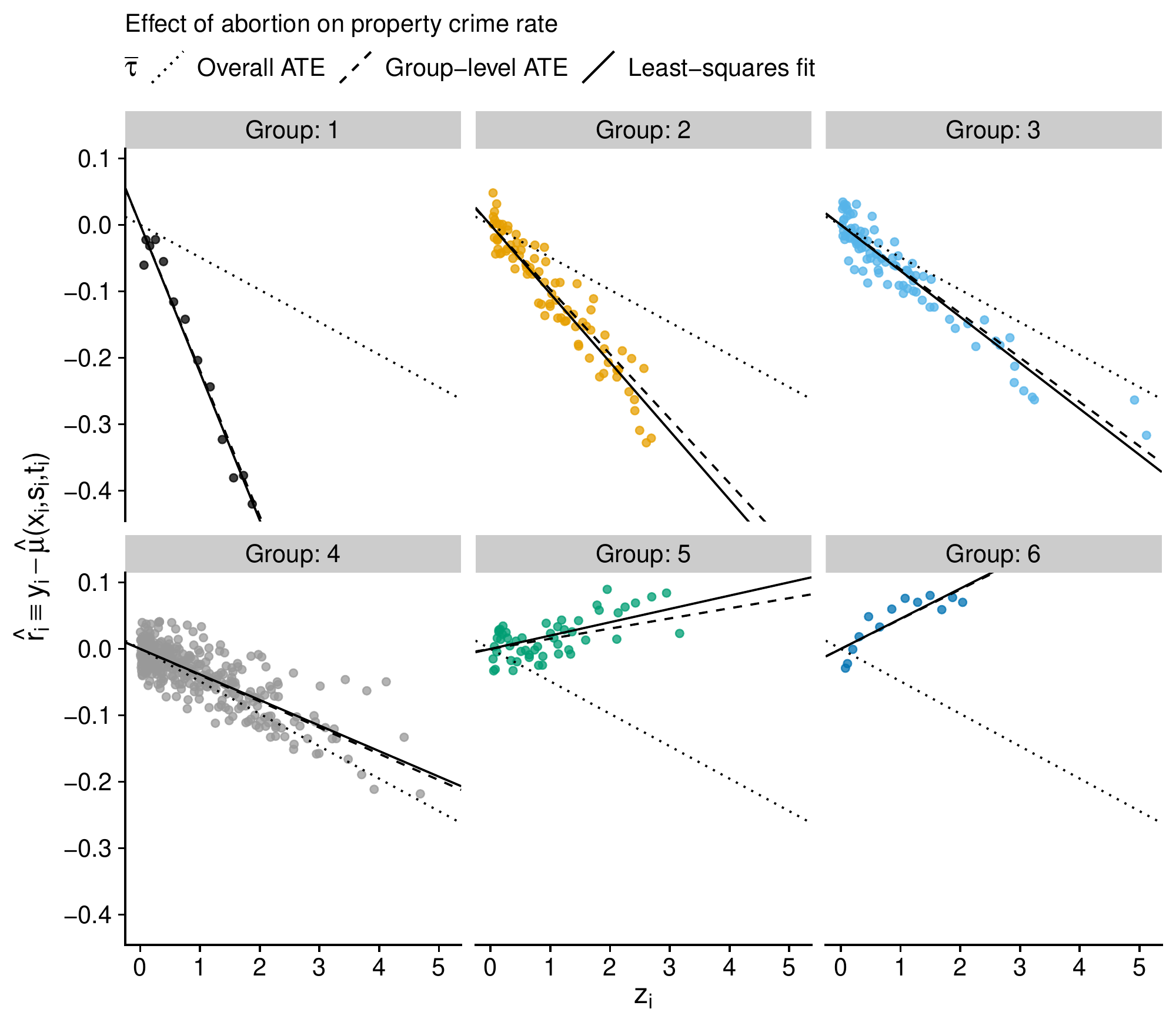}
  \caption{\label{fig:levitt-tau-clust-p} Diagnostics for the
    linearity assumption for the property crime data. }
\end{figure}

\begin{figure}[ht!]
  \centering
  \includegraphics[width=0.7\textwidth]{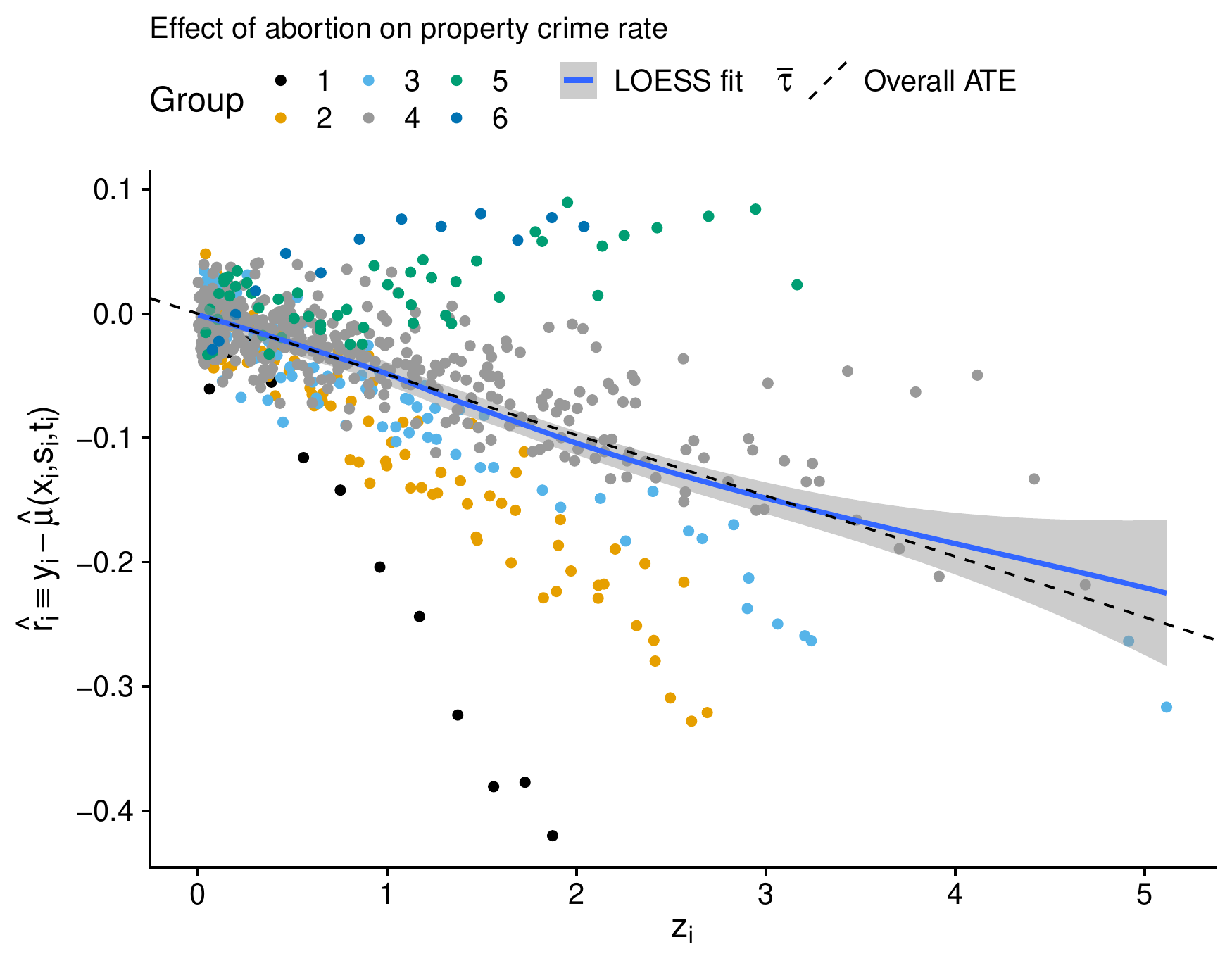}
  \caption{\label{fig:cluster-scatter-all-p} Plot of partial residuals
    $y_i - \mu(x_i)$ vs. treatment $z_i$ for every observation for the property crime data. }
\end{figure}

In the main text, we presented an analysis of the effect of abortion
on murder rate across the 48 contiguous United States for the years
1985--1997.  In line with \cite{Levitt}, we now present analyses of
the causal effect of abortion on violent crime
(Figures~\ref{fig:levitt-ate-v}, \ref{fig:levitt-state-v},
\ref{fig:proj-tau-mu-v}, \ref{fig:levitt-tau-clust-v}, and
\ref{fig:cluster-scatter-all-v} in Section~\ref{sec:v}) and property
crime (Figures~\ref{fig:levitt-ate-p}, \ref{fig:levitt-state-p},
\ref{fig:proj-tau-mu-p}, \ref{fig:levitt-tau-clust-p}, and
\ref{fig:cluster-scatter-all-p} in Section~\ref{sec:p}).

\section{Analysis of effect of abortion on violent crime}
\label{sec:v}

\section{Analysis of effect of abortion on property crime}
\label{sec:p}

\begin{singlespace}
\begin{raggedright}
\bibliography{main}

\end{raggedright}
\end{singlespace}


\end{document}